\documentclass[11pt]{article}
\usepackage[utf8]{inputenc}
\usepackage{amsthm,amsmath,amssymb,array,bm}
\usepackage{graphicx}
\usepackage{booktabs}
\usepackage{multirow}
\usepackage{lscape}
\usepackage{setspace}
\usepackage{hyperref}
\usepackage{authblk}
\usepackage{algorithm} 
\usepackage{algpseudocode} 
\usepackage{xcolor}

\usepackage{geometry}
\DeclareMathOperator{\Tr}{Tr}

\algnewcommand{\algorithmicand}{\textbf{ and }}
\algnewcommand{\algorithmicor}{\textbf{ or }}
\algnewcommand{\OR}{\algorithmicor}
\algnewcommand{\AND}{\algorithmicand}

\usepackage{natbib}
\title{\textbf{Covariance regression with random forests} \vspace{0.6cm}}
\author{Cansu Alakuş\thanks{Corresponding author. E-mail: \href{mailto:cansu.alakus@hec.ca}{cansu.alakus@hec.ca}}}
\author{Denis Larocque}
\author{Aur\'elie Labbe}
\affil{Department of Decision Sciences, HEC Montr\'eal, Montr\'eal, QC H3T 2A7, Canada} 
\date{}

\begin{document}

\maketitle
\begin{abstract}
Capturing the conditional covariances or correlations among the elements of a multivariate response vector based on covariates is important to various fields including neuroscience, epidemiology and biomedicine. We propose a new method called Covariance Regression with Random Forests (CovRegRF) to estimate the covariance matrix of a multivariate response given a set of covariates, using a random forest framework. Random forest trees are built with a splitting rule specially designed to maximize the difference between the sample covariance matrix estimates of the child nodes. We also propose a significance test for the partial effect of a subset of covariates. We evaluate the performance of the proposed method and significance test through a simulation study which shows that the proposed method provides accurate covariance matrix estimates and that the Type-1 error is well controlled. An application of the proposed method to thyroid disease data is also presented. \texttt{CovRegRF} is implemented in a freely available R package on CRAN.
\end{abstract}

\newpage

\section{Introduction} \label{sec:Introduction}

Most existing multivariate regression analyses focus on estimating the conditional mean of the response variable given its covariates. For example, in traditional regression analysis, the expectation of the response variables is related to linear combinations of covariates. While estimating the conditional covariances or correlations among multiple responses based on covariates is also important, it is a less studied problem. For example, functional brain connectivity focuses on the exploration of the co-occurrence of brain activity in different brain regions, and this co-variability can be explained as a function of covariates \citep{seiler_multivariate_2017}. As another example, human biomarkers such as glucose, cholesterol, iron, albumin, and so on, are important for biomedical research and the covariance of these biomarkers is influenced by age  \citep{le_goallec_age-dependent_2019}. In microbiome studies, the changes in the co-occurrence patterns among taxa with respect to the covariates have been studied \citep{levy_metabolic_2013, mcgregor_mdine_2020}. In tasks of cognitive and physical performance, a research question is whether the correlation between speed and accuracy is influenced by other covariates, such as sustained attention or age \citep{tu_cocoa_2022}. In neuroscience, the associations of functional criticality with intelligence can be affected by age \citep{jiang2021distance}. In all these examples, the main goal could be just to estimate the conditional covariance between multiple responses, e.g. for microbiome data, the goal is to estimate network changes with respect to a set of covariates. Another interesting application of estimating covariance matrices based on covariates is to verify the homoscedasticity assumption in classical multivariate regression. In cases where testing the effect of covariates on the covariability of the response variables leads to the rejection of the null hypothesis, conditional estimates of the covariance matrix can be used to have valid inference, for example to build multivariate confidence or prediction regions.

In general terms, let $\mathbf{Y}_{n \times q}$ be a matrix of $q$ response variables measured on $n$ observations, where $\mathbf{y}_i^\top$ represents the $i$th row of $\mathbf{Y}$. Similarly, let $\mathbf{X}_{n \times p}$ be a matrix of $p$ covariates available for all $n$ observations, where $\mathbf{x}_i^\top$ represents the $i$th row of $\mathbf{X}$. For an observation with covariates $\mathbf{x}_i$ and responses $\mathbf{y}_i$, the goal is to estimate the conditional covariance of the response variables ${\rm Cov}[\mathbf{y}_i|\mathbf{x}_i]\triangleq \Sigma_{\mathbf{x}_i}$, which is a measurable matrix function of covariates $\mathbf{x}_i$, and to analyze how this conditional covariance matrix varies with respect to the covariates. For this problem, \cite{yin_nonparametric_2010} use a kernel estimator to estimate the conditional covariance matrix for a single continuous covariate. However, it is not clear how to extend this approach to situations with multiple covariates. \cite{hoff_covariance_2012} propose a linear covariance regression model
\begin{equation*}
    \mathbf{y}_i = \left(\mathbf{A} + \gamma_i \mathbf{B}\right) \begin{pmatrix} 1\\\mathbf{x}_i \end{pmatrix} + \bm{\epsilon}_i,
\end{equation*}
where the mean and covariance of the multivariate response is parameterized as functions of covariates. This model can also be interpreted as a special random-effects model where $\mathbf{A}_{q\times (p+1)}$ and $\mathbf{B}_{q\times (p+1)}$ characterize the fixed and random parts of the model, respectively. The scalar $\gamma_i$ can be interpreted as an individual-level variability in addition to the random error $\bm{\epsilon}_i$. The rows of $\mathbf{B}$ indicate how much this additional variability affects $\mathbf{y}_i$. The vector $\bm{\epsilon}_i$ is of dimension $q \times 1$ and is assumed to be normally distributed. In this framework, they assume that $E[\gamma_i]=0$, $E[\bm{\epsilon}_i]=0$, $E[\gamma_i \bm{\epsilon}_i]=0$, $Var[\gamma_i]=1$, $Var[\bm{\epsilon}_i]=\mathbf{\Psi}$, leading to the following covariance matrix
\begin{equation*}
\mathbf{\Sigma}_{\mathbf{x}_i} = \mathbf{\Psi} + \mathbf{B} \begin{pmatrix} 1\\\mathbf{x}_i \end{pmatrix} \begin{pmatrix} 1\\\mathbf{x}_i \end{pmatrix}^\top \mathbf{B}^\top.
\end{equation*}
\cite{niu_joint_2019} illustrate an application of this model with a four-dimensional health outcome. \cite{fox_bayesian_2015} propose a Bayesian nonparametric model for covariance regression within a high-dimensional response context. Their approach relates the high-dimensional multivariate response set to a lower-dimensional subspace through covariate-dependent factor loadings obtained with a latent factor model. The conditional covariance matrix is a quadratic function of these factor loadings. The method is limited to data sets with smaller sample sizes. \cite{franks_reducing_2021} proposes a parametric Bayesian model for high-dimensional responses. In this model, the conditional covariance matrices vary with continuous covariates. \cite{zou_covariance_2017} propose another covariance regression model where the covariance matrix is linked to the linear combination of similarity matrices of covariates. \cite{zhao_covariate_2021} propose a covariance regression method called Covariate Assisted Principal Regression (CAPR). Unlike the other covariance regression methods described in this section, the CAPR aims to find a linear projection of the multivariate response data such that the covariates can best describe the data variation in the projected space. The model assumes that in the eigendecomposition of covariance matrices, all covariance matrices in the sample are diagonalized by the same orthogonal matrix which results in a restrictive covariance matrix form.

In this study, we propose a nonparametric covariance regression method for estimating the covariance matrix of a multivariate response given a set of covariates, using a random forest framework. The above-mentioned methods are very useful in modeling covariance matrix but compared to them the proposed method offers higher flexibility in estimating the covariance matrix given the set of covariates. For example, with the proposed method, we can estimate the conditional covariance matrix for a set of covariates including multiple continuous and categorical variables, and the proposed method can be used to capture complex interaction patterns with the set of covariates. Moreover, the proposed method is nonparametric and needs less computational time compared to the parametric models, and can be applied to data sets with larger sample sizes.

Random forest \citep{breiman_random_2001} is an ensemble tree-based algorithm involving many decision trees, and can also be seen as an adaptive nearest neighbour predictor \citep{hothorn_bagging_2004, lin_random_2006, moradian_l1_2017, moradian_survival_2019, roy_prediction_2020, tabib_non-parametric_2020, alakus_conditional_2021}. In the proposed random forest framework, we grow each tree with a splitting rule specially designed to maximize the difference in the sample covariance of $\mathbf{Y}$ between child nodes. For a new observation $\mathbf{y}^*$ with covariates $\mathbf{x}^*$, the proposed random forest finds the set of nearest neighbour observations among the out-of-bag (OOB) observations that are not used in the tree growing process. This set of nearest neighbour observations is then used to estimate the conditional covariance matrix of $\mathbf{y}^*$ given $\mathbf{x}^*$. In each tree built in the proposed random forest framework, the set of covariates is used to find subgroups of observations with similar conditional covariance matrices, assuming that they are related to conditional covariance matrices. We propose a hypothesis test to evaluate the effect of a subset of covariates on the estimated covariance matrices while controlling for the others. We investigate two particular cases, the global effect of the covariates and the partial effect of a single covariate. 

This paper is organized as follows. In Section \ref{sec:method}, we give the details of the proposed method, significance test and variable importance measure. The simulation study results for accuracy evaluation, global and partial effects of covariates, and variable importance are presented in Section \ref{sec:simulation}. We provide a real data example in Section \ref{sec:realdata}, and conclude with some remarks in Section \ref{sec:conclusion}.

\section{Method} \label{sec:method}

Let $\mathbf{\Sigma}_{\mathbf{x}_i}$ be the true conditional covariance matrix of $\mathbf{y}_i$ based on covariates $\mathbf{x}_i$, and $\mathbf{\Sigma}_{\mathbf{X}}$ be the collection of all conditional covariance matrices for $n$ observations, $\mathbf{\Sigma}_{\mathbf{X}}=\{ \mathbf{\Sigma}_{\mathbf{x}_i}:i=1,\dots,n\}$. Similarly, let $\mathbf{\hat \Sigma}_{\mathbf{x}_i}$ be the estimated conditional covariance matrix of $\mathbf{y}_i$ based on covariates $\mathbf{x}_i$, and $\mathbf{\hat \Sigma}_{\mathbf{X}}$ be the collection of all estimated conditional covariance matrices for $n$ observations, $\mathbf{\hat \Sigma}_{\mathbf{X}}=\{ \mathbf{\hat \Sigma}_{\mathbf{x}_i}:i=1,\dots,n\}$. In this section, we describe the proposed method in detail.

\subsection{Tree growing process and estimation of covariance matrices for new observations with random forests}

We aim to train a random forest with the set of covariates $\mathbf{X}$ to find subgroups of observations with similar covariance matrices of $\mathbf{Y}$, based on many unsupervised decision trees built with a specialized splitting criterion. The tree growing process follows the CART approach \citep{breiman_classification_1984}. The basic idea of the CART algorithm is to select the best split at each parent node among all possible splits, all evaluated with a selected splitting criterion, to obtain the purest child nodes. The algorithm evaluates all possible splits to determine the split variable and split point. Instead of considering all possible splits at each parent node, the best split search in random forests is confined to a randomly chosen subset of covariates that varies from node to node. The splitting process continues until all nodes are terminal.

Our goal is to obtain subgroups of observations with distinct covariance matrices. Hence, we propose a customized splitting rule that will seek to increase the difference in covariance matrices between two child nodes in the tree \citep{athey_generalized_2019, moradian_l1_2017, tabib_non-parametric_2020, alakus_conditional_2021}. We define $\mathbf{\Sigma}^L$ as the sample covariance matrix estimate of the left node as follows: 
\begin{equation*}
    \mathbf{\Sigma}^L = \frac{1}{n_L-1}\sum_{i \in t_L} \left(\mathbf{y}_i - \mathbf{\bar Y}_L \right)\left(\mathbf{y}_i - \mathbf{\bar Y}_L\right)^\top,
\end{equation*}
where $t_L$ is the set of indices of the observations in the left node, $n_L$ is the left node size and $\mathbf{\bar Y}_L = \frac{1}{n_L} \sum_{i \in t_L} \mathbf{y}_i$. The sample covariance matrix estimate of the right node, $\mathbf{\Sigma}^R$, is computed in the same way, where $n_R$ is the right node size.
The proposed splitting criterion is 
\begin{equation} \label{eq:split} 
\sqrt{n_Ln_R}*d(\mathbf{\Sigma}^L, \mathbf{\Sigma}^R),
\end{equation} 
where $d(\mathbf{\Sigma}^L, \mathbf{\Sigma}^R)$ is the Euclidean distance between the upper triangular part of the two matrices and computed as follows:
\begin{equation} \label{eq:eucdist} 
d\left(\mathbf{D}, \mathbf{E}\right) = \sqrt{\sum_{i=1}^{q}\sum_{j=i}^{q} \left(\mathbf{D}_{ij} - \mathbf{E}_{ij}\right)^2},
\end{equation}
where $\mathbf{D}_{q \times q}$ and $\mathbf{E}_{q \times q}$ are symmetric matrices. The best split among all possible splits is the one that maximizes \eqref{eq:split}.

The final covariance matrices are estimated based on the random forest. For a new observation, we use the nearest neighbour observations to estimate the final covariance matrix. The idea of finding the nearest neighbour observations, a concept very similar to the ‘nearest neighbour forest weights’ \citep{hothorn_bagging_2004, lin_random_2006}, was introduced in \cite{moradian_l1_2017} and later used in \cite{moradian_survival_2019, roy_prediction_2020, tabib_non-parametric_2020, alakus_conditional_2021}. \cite{roy_prediction_2020} called this set of observations the Bag of Observations for Prediction (BOP). 

For a new observation $\mathbf{x}^{*}$, we form the set of nearest neighbour observations with the out-of-bag (OOB) observations \citep{lu_unified_2021, alakus_r_2022}. We can define the $BOP_{oob}$ for a new observation as
\begin{equation*}
    BOP_{oob}(\mathbf{x}^{*}) = \bigcup\limits_{b=1}^{B} O_b(\mathbf{x}^{*}),
\end{equation*}
where $B$ is the number of trees and $O_b(\mathbf{x}^{*})$ is the set of OOB observations in the same terminal node as $\mathbf{x}^{*}$ in the $b$th tree. Each tree is built with a selected random sub-sample instead of a bootstrap sample, i.e. in-bag observations ($I_b$), which has 63.2 percent distinct observations from the original sample. The remaining training observations, namely $O_b$, are OOB observations for that tree and are not used to build the $b$th tree. $BOP_{oob}$ is slightly different than the nearest neighbour sets in the previous papers who use in-bag observations to form BOP. Since the OOB observations are not used in the tree building process, for the trees where they are OOB, they act as new observations. Therefore, OOB observations represent a new observation better than in-bag observations. Using OOB observations for neighbourhood construction is similar to the idea of honesty in the context of forests. An honest double-sample tree splits the training subsample into two parts: one part for tree growing and another part for estimating the desired response \citep{wager_estimation_2018}. We use the nearest neighbour construction idea to estimate the covariance matrices for the new observations. Algorithm \ref{alg:covest} describes how to estimate the covariance matrix with OOB observations for a new or training observation. After training the random forest with the specialized splitting criterion, for a new observation $\mathbf{x}^{*}$, we form $BOP_{oob}(\mathbf{x}^{*})$ and then we estimate the covariance matrix by computing the sample covariance matrix of the observations in $BOP_{oob}(\mathbf{x}^{*})$. See Supplementary figures \ref{fig:relativeacc} and \ref{fig:relativeaccglobal} in the Supplementary Material for the results of the simulation study comparing different ways of estimating the final covariance matrix.

\subsection{\texttt{nodesize} tuning} \label{subsec:nodesize}

The number of observations in the nodes decreases as we progress down the tree during the tree-building process. The \texttt{nodesize} parameter is the target average size for the terminal nodes. Lowering this parameter results in deeper trees, which means more splits until the terminal nodes. Tuning the \texttt{nodesize} parameter can potentially improve the prediction performance \citep{lin_random_2006}.

In typical supervised problems where the target is the observed true response, random forests search for the optimal level of the \texttt{nodesize} parameter by using out-of-bag (OOB) prediction errors computed using the true responses and OOB predictions. The \texttt{nodesize} value with the smallest OOB error is chosen. However, in our problem, the target is the conditional covariance matrix which is unknown. Therefore, we propose a heuristic method for tuning the \texttt{nodesize} parameter. For \texttt{nodesize} tuning, we use the OOB covariance matrix estimates, as described in Algorithm \ref{alg:covest}.

\begin{algorithm}[htbp]
	\caption{Estimation of covariance matrix for a new or training observation} 
	\begin{algorithmic}[1]
	\Require A forest built with the proposed method
	\State $BOP_{oob}(\mathbf{x}_i) = \emptyset$
        \For {b=1,...,B}
            \If {$\mathbf{x}_i$ is a new observation \OR ($\mathbf{x}_i$ is a training observation \AND $\mathbf{x}_i \in O_b$)}
                \State Find the terminal node of $\mathbf{x}_i$ at tree $b$, say $d$
                \State $BOP_{oob}(\mathbf{x}_i) = BOP_{oob}(\mathbf{x}_i) \cup O_b^d(\mathbf{x}_i)$ (where $O_b^d(\mathbf{x}_i)$ is the set of OOB observations in the same terminal node $d$ as $\mathbf{x}_i$, excluding $\mathbf{x}_i$ itself when $\mathbf{x}_i$ is a training observation)
            \EndIf
        \EndFor
        \State Compute sample covariance matrix with the observations in $BOP_{oob}(\mathbf{x}_i)$
	\end{algorithmic} 
	\label{alg:covest}
\end{algorithm}

The general idea of the \texttt{nodesize} tuning method is to find the \texttt{nodesize} level where the average difference between OOB covariance matrix predictions at two consecutive \texttt{nodesize} levels is the smallest among the set of \texttt{nodesize} values. We first train separate random forests for a set of \texttt{nodesize} values (see the Parameter settings section in simulation study). Then, we compute the OOB covariance matrix estimates as described in Algorithm \ref{alg:covest} for each random forest. Define $MAD\left(\mathbf{D},\mathbf{E}\right)= \frac{2}{q(q+1)} \sum_{i=1}^{q}\sum_{j=i}^{q} | \mathbf{D}_{ij} - \mathbf{E}_{ij} |$. Let $\mathbf{\hat \Sigma}^s_{\mathbf{x}_{i}}$ be the estimated covariance matrix for observation $i$ when \texttt{nodesize}$=s$. Let $s(1) < \ldots < s(M)$ be a set of increasing node sizes. For $j=\{1,\ldots M-1\}$, let $$MAD_j=\frac{1}{n}\sum_{i=1}^{n} MAD \left( \mathbf{\hat \Sigma}_{\mathbf{x}_{i}}^{s(j)},\mathbf{\hat \Sigma}_{\mathbf{x}_{i}}^{s(j+1)} \right).$$
Then we select $s(j)$ that corresponds to the value $j$ for which $MAD_j$ is the minimum among $\{MAD_1,\ldots, MAD_M\}$. See Section \ref{suppsec:2_1} of the Supplementary Material for the results of a \texttt{nodesize} tuning experiment and the illustration of the process with an example.

When a node sample size $n_d$ is smaller than the number of responses $q$, the sample covariance matrix becomes highly variable. In fact, if $n_d - 1 < q$, the estimate is singular and hence non-invertible. Therefore, the tuning set of \texttt{nodesize} levels should be larger than $q$. In fact, we need more than $q$ distinct values, so we use sub-sampling instead of bootstrap resampling for each tree building step of the proposed method to guarantee distinctness, assuming the observations in the original sample are distinct.

\subsection{Significance test}
\label{subsec:significance}

The proposed method uses covariates to find groups of observations with similar covariance matrices with the assumption that the set of covariates is important to distinguish between these covariance matrices. However, some (or all) covariates might not be relevant. In this paper, we propose a hypothesis test to evaluate the effect of a subset of covariates on the covariance matrix estimates, while controlling for the other covariates. 

If a subset of covariates has an effect on the covariance matrix estimates obtained with the proposed method, then the conditional covariance matrix estimates given all covariates should be significantly different from the conditional covariance matrix estimates given the controlling set of covariates. We propose a hypothesis test to evaluate the effect of a subset of covariates on the covariance matrix estimates for the null hypothesis 
\begin{equation} \label{eq:null0}
    H_0 : \mathbf{\Sigma}_\mathbf{X} = \mathbf{\Sigma}_{\mathbf{X}^c},
\end{equation}
where $\mathbf{\Sigma}_\mathbf{X}$ is the conditional covariance matrix of $\mathbf{Y}$ given all $X$ variables, and $\mathbf{\Sigma}_{\mathbf{X}^c}$ is the conditional covariance matrix of $\mathbf{Y}$ given only the set of controlling $X$ variables. The proposed significance test is described in Algorithm \ref{alg:significance}. After computing the covariance matrix estimates for all covariates and control variables only, we compute the test statistic with 
\begin{equation} \label{eq:test}
    T = \frac{1}{n} \sum_{i=1}^{n}{d \big(\mathbf{\hat \Sigma}_{\mathbf{x}_i}, \mathbf{\hat \Sigma}_{\mathbf{x}^c_i}\big)},
\end{equation}
where $d(.,.)$ is computed as \eqref{eq:eucdist}. The test statistic specifies how much the covariance matrix estimates given all covariates differ from the estimates given only the controlling set of covariates. As $T$ becomes larger, we have more evidence against $H_0$. 

We conduct a permutation test under the null hypothesis \eqref{eq:null0} by randomly permuting rows of $\mathbf{X}$. Let $R$ be the total number of permutations and $T_r$ be the global test statistic \eqref{eq:test} computed for the $r$th permuted $\mathbf{X}$. We estimate the test $p$-value with 
\begin{equation} \label{eq:pvalue}
    p = \frac{1}{R} \sum_{r=1}^{R}{I(T_r > T)},
\end{equation} 
and we reject the null hypothesis \eqref{eq:null0} at a pre-specified level $\alpha$ if the $p$-value is less than  $\alpha$. 

\begin{algorithm}[]
	\caption{Permutation test for a subset of covariates effect} 
	\begin{algorithmic}[1]
	\State Train RF with $\mathbf{X}$ and $\mathbf{Y}$, estimate covariance matrices as described in Algorithm \ref{alg:covest}, say $\mathbf{\hat \Sigma}^{}_{\mathbf{x}_i}$ $\forall i=\{1,\ldots,n\}$
	\State Train RF with $\mathbf{X}^{c}$ and $\mathbf{Y}$, and estimate covariance matrices as described in Algorithm \ref{alg:covest}, say $\mathbf{\hat \Sigma}^{c}_{\mathbf{x}_i}$ $\forall i=\{1,\ldots,n\}$
	\State Compute test statistic with $T = \frac{1}{n} \sum_{i=1}^{n}{d \left(\mathbf{\hat \Sigma}^{}_{\mathbf{x}_i}, \mathbf{\hat \Sigma}^{c}_{\mathbf{x}_i}\right)}$, where $d(.,.)$ is computed as \eqref{eq:eucdist}
	\For {$r = 1:R$}
	    \State Permute rows of $\mathbf{X}$, say $\mathbf{X}_{r}$
	    \State Train RF with $\mathbf{X}_{r}$ and $\mathbf{Y}$
	    \State Estimate covariance matrices as described in Algorithm \ref{alg:covest}, say $\mathbf{\hat \Sigma}^{'}_{\mathbf{x}_i}$ $\forall i=\{1,\ldots,n\}$
	    \State Train RF with $\mathbf{X}^{c}_{r}$ and $\mathbf{Y}$
	    \State Estimate covariance matrices as described in Algorithm \ref{alg:covest}, say $\mathbf{\hat \Sigma}^{c'}_{\mathbf{x}_i}$ $\forall i=\{1,\ldots,n\}$
	    \State Compute test statistic with $T_r = \frac{1}{n} \sum_{i=1}^{n}{d \left(\mathbf{\hat \Sigma}^{'}_{\mathbf{x}_i}, \mathbf{\hat \Sigma}^{c'}_{\mathbf{x}_i}\right)}$
    \EndFor
    \State Approximate the permutation $p$-value with $p = \frac{1}{R} \sum_{r=1}^{R}{I(T_r > T)}$
    \State Reject the null hypothesis when $p < \alpha$. Otherwise, do not reject the null hypothesis.
	\end{algorithmic} 
	\label{alg:significance}
\end{algorithm}

In the significance test described above, we need to apply the proposed method many times: for the original data with \textit{(i)} all covariates and \textit{(ii)} the set of control covariates, and at each permutation for the permuted data with \textit{(iii)} all covariates and \textit{(iv)} the set of control covariates. The proposed method applies a \texttt{nodesize} tuning as described in the previous section. Since tuning the \texttt{nodesize} parameter can be computationally demanding, we tune the \texttt{nodesize} for the original data with all covariates and with the set of control covariates only and use those tuned values for their corresponding permutation steps.

The proposed significance test has two particular cases of interest. The first is to evaluate the global effect of the covariates on the conditional covariance estimates. If $\mathbf{X}$ has a global effect on the covariance matrix estimates obtained with the proposed method, then the conditional estimates $\mathbf{\Sigma}_\mathbf{X}$ should be significantly different from the unconditional covariance matrix estimate $\mathbf{\Sigma}_{root}$ which is computed as the sample covariance matrix of $\mathbf{Y}$. The null hypothesis \eqref{eq:null0} becomes
\begin{equation} \label{eq:null1}
    H_0 : \mathbf{\Sigma}_\mathbf{X} = \mathbf{\Sigma}_{root}.
\end{equation}
See the Supplementary Algorithm \ref{alg:global} in the Supplementary Material for the details of the global significance test. The second case is to evaluate the effect of a single covariate when the other covariates are in the model. In that particular case, the null hypothesis \eqref{eq:null0} remains. The only difference between the global and partial significance tests is the number of forests we need to train. In the partial significance test, we need to train two random forests per sample, one for all covariates and one for the controlling variables, which makes a total $2R+2$ random forests. However, when we test for the global effect, we need to train only one random forest per sample (in total $R+1$ random forests) since  we do not need to build a random forest for the root node.

\subsection{Variable importance} \label{subsec:variableimportance}

For traditional regression tree problems, we can get the variable importance (VIMP) measures by computing the average change in prediction accuracy using the OOB samples. However, the covariance regression problem does not have an observed target. We can compute the VIMP measures by using the \textit{fit-the-fit} approach which has been applied to enhance interpretability of the covariates on the response \citep{lee_causal_2020, alakus_conditional_2021, spanbauer_nonparametric_2021, bargagli-stoffi_assessing_2021, bargagli-stoffi_heterogeneous_2022, meid_machine_2022}. In the univariate response case, we get the importance measures by fitting a regression forest to re-predict the predicted values. However, in covariance regression, we have a predicted covariance matrix for each observation and not a single value. Therefore, we use a multivariate splitting rule based on the Mahalanobis distance \citep{ishwaran_randomforestsrc_2021} to re-predict the predicted covariance matrices. We begin by applying the proposed method using the original covariates and responses and estimate the covariance matrices as described in Algorithm \ref{alg:covest}. Next, we train a random forest with the original covariates and the vector of upper-triangular estimated covariance matrix elements as a multivariate response. VIMP measures are obtained from this random forest. Covariates with higher VIMP measures indicate higher importance for the estimation of covariance matrices. The proposed VIMP computation is described in Supplementary Algorithm \ref{alg:vimp} in the Supplementary Material.

\subsection{Software}

We have developed an R package called \texttt{CovRegRF}. We used the custom splitting feature of the \texttt{randomForestSRC} package \citep{R-randomForestSRC} to implement our specially designed splitting criterion in the tree building process. The package is available on CRAN, \url{https://CRAN.R-project.org/package=CovRegRF}.

\section{Simulations} \label{sec:simulation}

In this section, we perform a simulation study to demonstrate the performance of the proposed method, validate the proposed significance test with two particular cases—global and partial significance tests—and evaluate the variable importance estimations of the covariates.

\subsection{Data generating process}

We carry out a simulation study using four Data Generating Processes (DGPs). The details of the DGPs are given in Section \ref{suppsec:dgp} of the Supplementary Material. The first two DGPs are variations of the first simulated data set used in \cite{hoff_covariance_2012}. Both DGPs include one covariate and two response variables. The covariate $x$ is generated uniformly on $[-1, 1]$. In DGP1, the covariance matrix for the observation $x_i$ is $\mathbf{\Sigma}_{\mathbf{x}_i} = \mathbf{\Psi} + \mathbf{B} \mathbf{x}_i^{} \mathbf{x}_i^\top \mathbf{B}^\top$ where $\mathbf{x}_i^\top=(1, x_i)^\top$. DGP2 is similar to DGP1, except that we add a quadratic term to the covariance matrix equation such as $\mathbf{\Sigma}_{\mathbf{x_i}} = \mathbf{\Psi} + \mathbf{B} \mathbf{\dot{x}}_i^{} \mathbf{\dot{x}}_i^\top \mathbf{B}^\top$ where $\mathbf{\dot{x}}_i^\top=(1, (x_i + x_i^2))^\top$.

In DGP3, the vector of covariates includes seven independent variables generated from the standard normal distribution. For the covariance structure, we use an AR(1) structure with heterogeneous variances. The correlations are generated with all seven covariates according to a tree model with a depth of three and eight terminal nodes. The variances are functions of the generated correlations. In DGP4, the covariance matrix has a compound symmetry structure with heterogeneous variances. Both variances and correlations are functions of covariates. The covariates are generated from the standard normal distribution. The correlations are generated with a logit model and the variances are functions of these generated correlations. The number of covariates and response variables varies depending on the simulation settings. For all DGPs, after generating $\mathbf{\Sigma}_{\mathbf{x}_i}$, $\mathbf{y}_i$ is generated from a multivariate normal distribution $N(\mathbf{0},\mathbf{\Sigma}_{\mathbf{x}_i})$.

\subsection{Simulation design}

\subsubsection{Accuracy evaluation}

We perform a simulation study based on the four DGPs described above to evaluate the accuracy of the proposed method for estimating the covariance matrices. For DGP3 and DGP4, we consider five response variables. For each DGP, we use several values of the training sample size $n_{train}=\{50,100,200,500,1000\}$, which generates a total of 20 settings (4 DGPs $\times$ 5 training sample sizes). We repeat each setting 100 times. In each run of the simulations, we generate an independent test set of new observations with $n_{test} = 1000$. 

We evaluate the performance of the covariance matrix estimates using the mean absolute errors (MAE) computed for both the estimated correlations and standard deviations separately. For the estimated correlations, we compute the MAE between the upper triangular (off-diagonal) matrices of the true and estimated correlations over all observations as follows:
\begin{equation*}
    MAE^{cor}(\mathbf{\hat  C}_\mathbf{X},\mathbf{C}_\mathbf{X}) = \frac{2}{q(q-1)n_{test}}\sum_{i=1}^{n_{test}} \sum_{j=1}^{q} \sum_{k=j+1}^q |\hat \rho_{ijk} - \rho_{ijk}|,
\end{equation*}
where $\mathbf{C}_\mathbf{X}$ and $\mathbf{\hat C}_\mathbf{X}$ are the collection of all correlation matrices corresponding to $\Sigma_\mathbf{X}$ and $\hat \Sigma_\mathbf{X}$, respectively. The values $\rho_{ijk}$ and $\hat \rho_{ijk}$ represent the correlations in row $j$ and column $k$ of $\mathbf{C}_{\mathbf{x}_i}$ and $\mathbf{\hat C}_{\mathbf{x}_i}$, respectively. 

For the estimated standard deviations, we compute the normalized MAE between the true and estimated standard deviations over all observations as follows:
\begin{equation*}
    MAE^{sd}(\mathbf{\hat \Sigma}_{\mathbf{X}}, \mathbf{\Sigma}_{\mathbf{X}}) = \frac{1}{qn_{test}}\sum_{i=1}^{n_{test}} \sum_{j=1}^{q} \Bigg|\frac{\hat \sigma_{ij} - \sigma_{ij}}{\sigma_{ij}}\Bigg|.
\end{equation*}
The values $\sigma^2_{ij}$ and $\hat \sigma^2_{ij}$ represent the $j$th diagonal element of $\mathbf{\Sigma}_{\mathbf{x}_i}$ and $\mathbf{\hat \Sigma}_{\mathbf{x}_i}$, respectively. 

Smaller values of $MAE^{cor}$ and $MAE^{sd}$ indicate better performance. We compare our proposed method with the original Gaussian-based covariance regression model \texttt{covreg} developed in \cite{hoff_covariance_2012} which was presented in the Introduction. This method is currently available in the \texttt{covreg} R package \citep{R-covreg}. Moreover, as a simple benchmark method, we compute the sample covariance matrix without covariates, which is then used as the covariance matrix estimate for all new observations from the test set. 

\subsubsection{Variable importance}

For the variable importance evaluation simulations, we use DGP3 and DGP4 in which we add five noise variables $X$ to the covariates set. As above, we consider several values for the training sample sizes $n_{train}=\{50, 100, 200, 500, 1000\}$, for a total of 10 scenarios studied. We examine whether the estimated VIMP measures tend to rank the important variables first. The variable with the highest VIMP measure has a rank of 1. For each scenario, we compute the average rank for the important variables group and for the noise variables group. 

\subsubsection{Evaluating the power of the global significance test}

We studied four scenarios to evaluate the global effect of the covariates, two of which are under the null hypothesis \eqref{eq:null1} and the other two under the alternative hypothesis. We generate the data sets for these scenarios as follows:
\begin{enumerate}
    \item $H_0$ (case 1): we generate 5 $Y$ with a constant population covariance matrix and 10 $X$ variables which are all independent following a standard normal distribution. In this case, the covariance of $Y$ is independent of $X$ and we are therefore under the null hypothesis. 
    \item $H_0$ (case 2): we first generate 7 $X$ and 5 $Y$ under DGP3. Then, we replace the $\mathbf{X}$ matrix with 10 independent $X$ variables generated from a standard normal distribution. In this case, the covariance of $\mathbf{Y}$ varies with some of the $X$ variables but those $X$ variables are not available in the training set. Therefore, we are again under the null hypothesis. 
    \item $H_1$ (without noise): we generate 7 $X$ and 5 $Y$ under DGP3, and the covariates are available in the training set. In this case, the covariance of $\mathbf{Y}$ varies with all $X$ variables.
    \item $H_1$ (with noise): we generate 7 $X$ and 5 $Y$ under DGP3 and we add 3 independent $X$ variables to the covariates' training set. In this case, the covariance of $\mathbf{Y}$ varies with some of the $X$ variables but not all.
\end{enumerate}

\subsubsection{Evaluating the power of the partial significance test}

We can consider three scenarios to evaluate the effect of a single covariate, where one is under the null hypothesis \eqref{eq:null0} and the other two under the alternative hypothesis. We generate the data sets for these scenarios as follows:
\begin{enumerate}
    \item $H_0$: We first generate 2 $X$ and 5 $Y$ with DGP4 and we add 1 independent $X$ variable to the covariates' training set. In this case, the covariance of $\mathbf{Y}$ varies only with the first two $X$ variables. The control set of variables is $\{X_1, X_2\}$ and we evaluate the effect of the $X_3$ variable. Therefore, we are under the null hypothesis. 
    \item $H_1 (weakest)$: We generate 3 $X$ and 5 $Y$ with DGP4. In this case, the covariance of $\mathbf{Y}$ varies with all $X$ variables. The control set of variables is $\{X_1, X_2\}$ and we evaluate the effect of $X_3$, which has the weakest effect on the covariance matrix.
    \item $H_1 (strongest)$: We generate 3 $X$ and 5 $Y$ with DGP4. In this case, the covariance of $\mathbf{Y}$ again varies with all $X$ variables. But now the control set of variables is $\{X_2, X_3\}$ and we evaluate the effect of $X_1$, which has the strongest effect on the covariance matrix.
\end{enumerate}

For both the global and partial significance test simulations, we use training sample sizes of $n_{train}=\{50,100,200,300,500\}$. The number of permutations and the number of replications for each scenario are set to 500. We estimate the type-1 error as the proportion of rejection in the scenarios simulated under $H_0$ and the power as the proportion of rejection in the scenarios simulated under $H_1$. We estimate a $p$-value for each replication and we reject the null hypothesis if the $p$-value is less than the significance level $\alpha=0.05$. Finally, we compute the proportion of rejection over 500 replications.

\subsection{Parameter settings}

For the simulations, we use the following parameters for the proposed method. We set the number of trees to 1000. Letting $p$ be the number of covariates, then the number of covariates to randomly split at each node, \texttt{mtry}, is set to $\lceil p/3 \rceil$. The number of random splits for splitting a covariate at each node, \texttt{nsplit}, is set to $\max \{n_{train}/50, 10\}$. We tune the \texttt{nodesize} parameter with the set of \texttt{nodesize}$=\{[sampsize \times (2^{-1},2^{-2},2^{-3},\ldots)]>q\}$ where $q$ is the number of responses and \texttt{sampsize}$=0.632n_{train}$. In each replication, \texttt{covreg} is run in four independent chains for 8000 iterations, with the first half taken as burn-in.

\subsection{Results}

\subsubsection{Accuracy evaluation}

\begin{figure}
    \centerline{\includegraphics[width=0.75\textwidth]{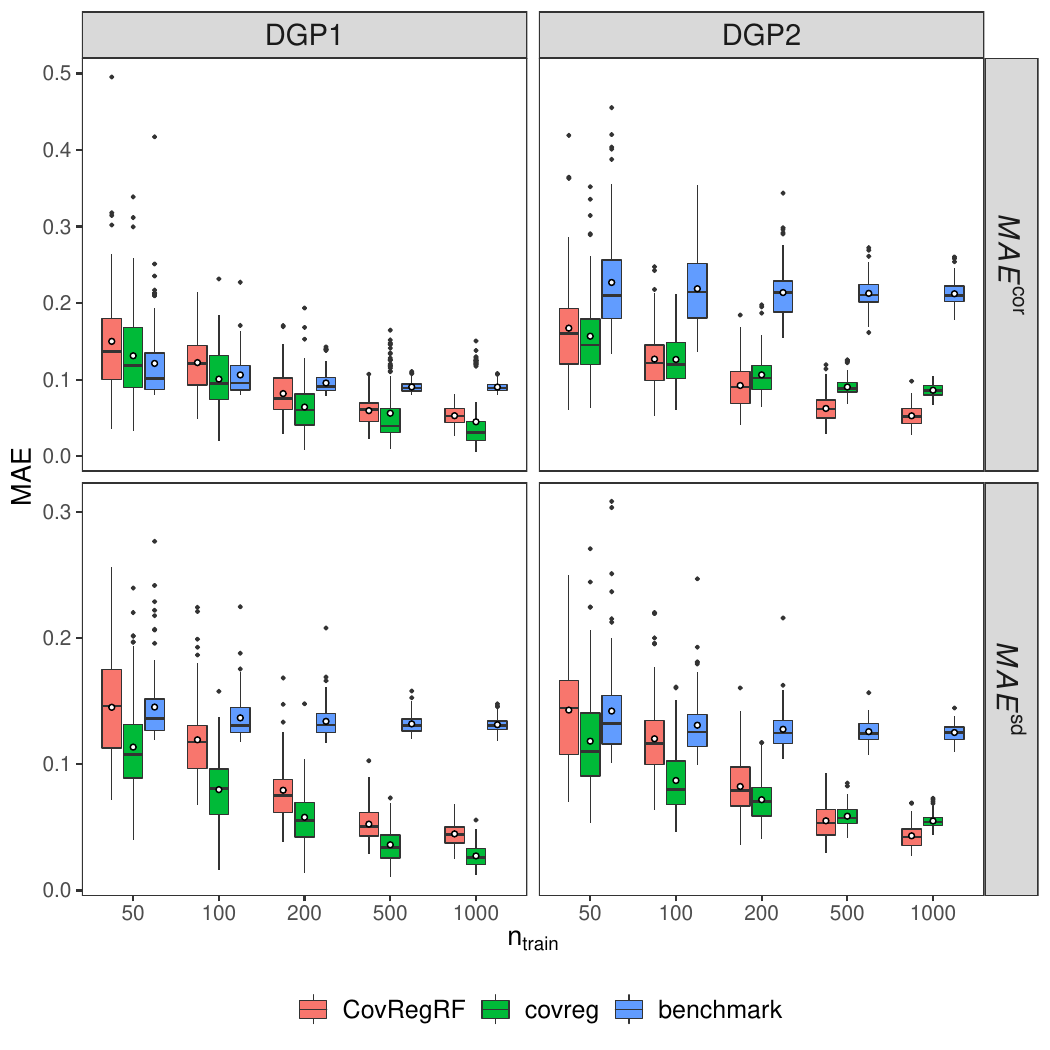}}
    \caption{Accuracy evaluation results for DGP1 and DGP2. Smaller values of $MAE^{cor}$ and $MAE^{sd}$ are better.}
    \label{fig:accuracy1}
\end{figure}

\begin{figure}
    \centerline{\includegraphics[width=0.75\textwidth]{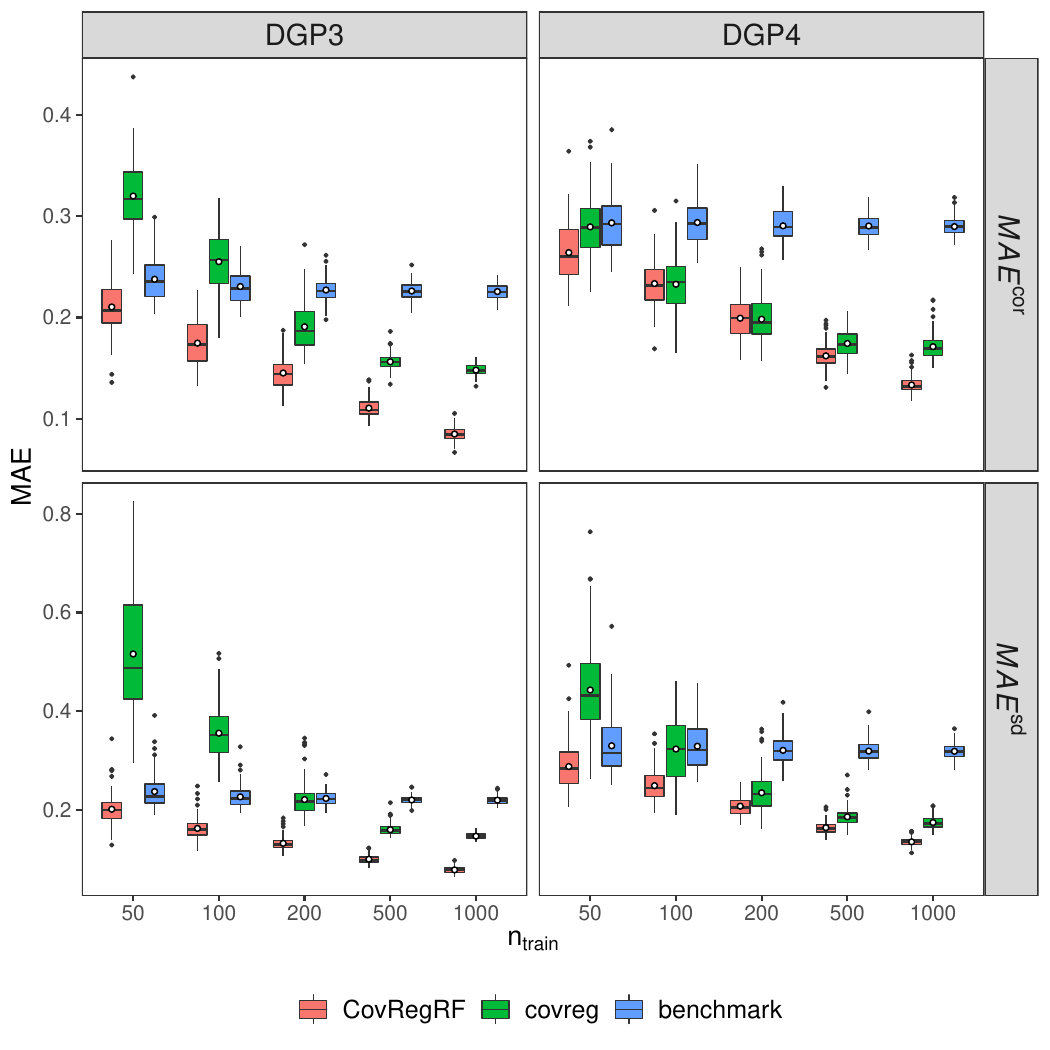}}
    \caption{Accuracy evaluation results for DGP3 and DGP4. Smaller values of $MAE^{cor}$ and $MAE^{sd}$ are better.}
    \label{fig:accuracy2}
\end{figure}

Figures \ref{fig:accuracy1} and \ref{fig:accuracy2} present the accuracy results for 100 repetitions. For each method, we can see the change in $MAE^{cor}$ and $MAE^{sd}$ computed for 100 repetitions with an increasing training sample size. As demonstrated in Figure \ref{fig:accuracy1}, for DGP1 and DGP2 when $n_{train}=50$, the proposed method and \texttt{covreg} both have a similar  performance with respect to the correlation estimation, with a slight advantage for \texttt{covreg}. For DGP1, \texttt{covreg} performs better for both the correlation and standard deviation compared to the proposed method as the sample size increases. This is expected since DGP1 is generated exactly under the \texttt{covreg} model. However, the proposed method still remains competitive. For DGP2, in which  a quadratic term is added, the proposed method performs better for the correlation  than \texttt{covreg} with increasing sample size. \texttt{covreg} shows better standard deviation estimation performance for  smaller sample sizes, but after $n_{train}=500$ the proposed method performs slightly better. As demonstrated in Figure \ref{fig:accuracy2}, for DGP3, the proposed method shows a significantly smaller $MAE^{cor}$ and $MAE^{sd}$ than \texttt{covreg} for all sample sizes. Moreover, for the smaller sample sizes, the proposed method has considerably lower variance in MAE. For DGP4, both methods improve with increasing sample size, but the proposed method shows smaller or equal MAEs for both correlation and standard deviation estimations. For DGP3 and DGP4, these results are expected, since the proposed method can capture a nonlinear effect. Supplementary figures \ref{fig:acc1} and \ref{fig:acc2} in the Supplementary Material present the difference in MAE between the proposed method and \texttt{covreg}. Moreover, we evaluate the accuracy with Stein's loss which is the Kullback–Leibler divergence between the estimated and true covariance matrices. The conclusions remain the same. See Supplementary Figure \ref{fig:acc3} in the Supplementary Material.

For the \texttt{nodesize} tuning, we compare the accuracy results for different levels of \texttt{nodesize} along with the proposed tuning method. Supplementary figures \ref{fig:nodesize1} and \ref{fig:nodesize2} in the Supplementary Material present the MAE results for all DGPs which show that the tuning method works well. 

\subsubsection{Variable importance} \label{subsec:vimpresult}

Supplementary Figure \ref{fig:vimp} in the Supplementary Material presents the average ranks of the VIMP measures for both the important and noise sets of variables for DGP3 and DGP4. In all scenarios, the important variables have smaller average ranks than noise variables. As the sample size increases, the difference between the average ranks of important and noise variables increases, as expected.

\subsubsection{Global significance test}

\begin{figure}
    \centerline{\includegraphics[width=0.8\textwidth]{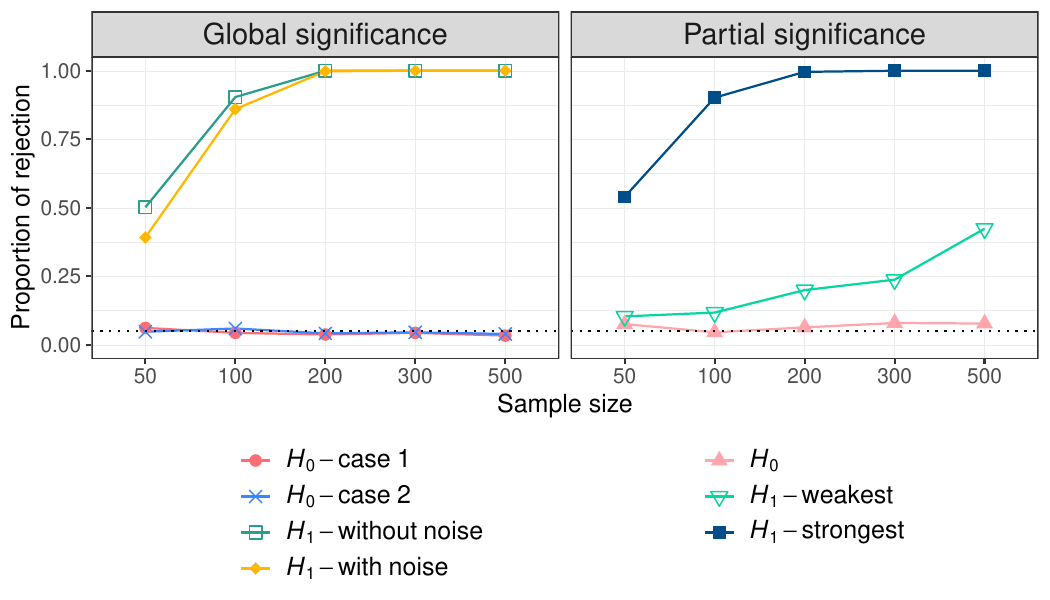}}
    \caption{Significance test results. The left and right plots present the results for global and partial significance tests, respectively. The proportion of rejection corresponds to the type-1 error for $H_0$ scenarios, and power for $H_1$ scenarios. The dotted line represents the significance level of $\alpha=0.05$.}
    \label{fig:sig}
\end{figure}

The left plot in Figure \ref{fig:sig} presents the estimated type-1 error and power for different training sample sizes for the two $H_0$ scenarios and two $H_1$ scenarios, respectively. We expect the type-1 error to be close to the significance level ($\alpha=0.05$) and we can see that it is well controlled in both cases studied. In both $H_1$ scenarios, the power increases with the sample size. When the sample size is small, adding noise covariates slightly decreases the power, but this effect disappears as the sample size increases. 

\subsubsection{Partial significance test}

The right plot in Figure \ref{fig:sig} presents the estimated type-1 error and power for different training sample sizes for the $H_0$ scenario and two $H_1$ scenarios, respectively. As can be seen from the $H_0$ line, the type-1 error is close to the significance level ($\alpha=0.05$). In both $H_1$ scenarios, the power increases with the sample size as expected. However, the power is much smaller when one tests the weakest covariate compared to the strongest covariate.

\section{Real data example} \label{sec:realdata}

Thyroid hormone, the collective name for two hormones, is widely known for regulating several body processes, including growth and metabolism \citep{yen_physiological_2001, shahid_physiology_2022}. The main hormones produced by the thyroid gland are triiodothyronine (T3) and thyroxine (T4). The synthesis and secretion of these hormones are primarily regulated by thyroid stimulating hormone (TSH), which is produced by the pituitary gland. Primary hypothyroidism is a condition that occurs when the thyroid gland is underactive and the thyroid hormone produced is insufficient to meet the body's requirements, which leads to an increase of TSH. Contrarily, when the thyroid gland produces levels of thyroid hormones that are too high, leading to decreased levels of TSH, the resulting condition is hyperthyroidism. 

Serum levels of the thyroid hormones and TSH are used to evaluate subjects' thyroid function status and to identify subjects with a thyroid dysfunction. Therefore, establishing reference intervals for these hormones is critical in the diagnosis of thyroid dysfunction. However, reference ranges are affected by age and sex \citep{kapelari_pediatric_2008, aggarwal_thyroid_2013, biondi_normal_2013, strich_ft3_2017, park_age-and_2018}. Furthermore, there is a relationship between TSH and thyroid hormone, and the effects of age and sex on this relationship have not been well described \citep{hadlow_relationship_2013, lee_causal_2020}. Serum levels of these hormones are also affected by the subject's diagnosis, i.e. hormone levels would be within the reference ranges for normal subjects and out of range for subjects with thyroid dysfunction. The conditional mean of these hormones based on the covariates is studied in the literature, but to our knowledge, no study has yet explicitly investigated the effect of covariates on the conditional covariance matrix of these hormones. Hence, our contribution is to study the effect of age, sex and diagnosis on the covariance matrix of the thyroid hormones and TSH. 

\begin{figure}[!h]
    \centerline{\includegraphics[width=0.85\textwidth]{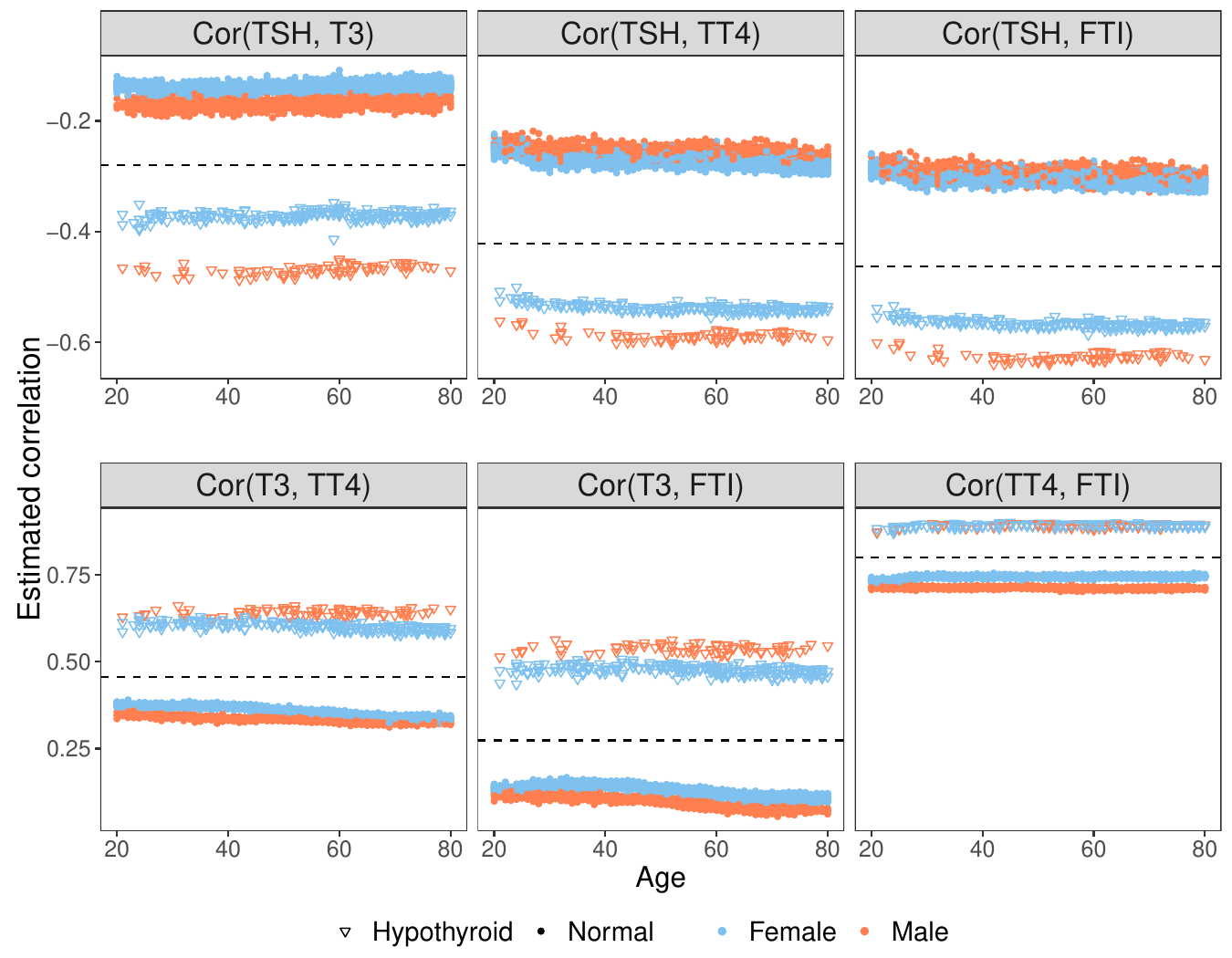}}
    \caption{Estimated correlations between the four hormones as a function of age, sex and diagnosis. Dashed lines represent the sample correlations computed using the whole sample.}
    \label{fig:thyroid}
\end{figure}

In this study, we investigate the thyroid disease data set from the UCI machine learning repository \citep{UCI}. This data set originally included 9172 subjects and 30 variables including age, sex, hormone levels and diagnosis. Following the exclusion criteria applied in \cite{hadlow_relationship_2013} and \cite{strich_ft3_2017}, we exclude pregnant women, subjects who have euthyroid sick syndrome (ESS), goitre, hypopituitarism or tumour, subjects who use antithyroid medication, thyroxine or lithium, who receive I131 treatment, or who have had thyroid surgery. The subjects have different diagnoses including hypothyroidism and hyperthyroidism, as well as normal subjects. Since the sample size of hyperthyroidism subjects is small, we exclude them from the analysis. We also exclude the very young and very old subjects, since there are only a few subjects on the extremes. The remaining data set consists of 324 hypothyroidism and 2951 normal subjects ($n=3275$) between 20 and 80 years of age (2021 females/1254 males). We want to estimate the covariance matrix of four thyroid-related hormones—TSH, T3, TT4 (total T4) and FTI (free thyroxine index/free T4)—based on covariates and investigate how the relationship between these hormones varies with the covariates. We apply the proposed method with the covariates age, sex and diagnosis to estimate the covariance matrix of the four hormones. We first perform the significance test with 500 permutations to evaluate the global effect of the three covariates. The estimated \textit{p}-value with \eqref{eq:pvalue} is 0 and we reject the null hypothesis \eqref{eq:null1}, which indicates that the conditional covariance matrices vary significantly with the set of covariates. Next, we apply the proposed method and obtain the covariance matrix estimates. We analyze the correlations between hormones as a function of covariates, and as shown in Figure \ref{fig:thyroid}, age seems not to have much effect on the estimated correlations. We also compute the variable importance measures, and age (0.001) is found to be the least important variable where diagnosis (1.000) is the most important variable, followed by sex (0.011). Therefore, we apply the significance test to evaluate the effect of age on covariance matrices while controlling for sex and diagnosis. Using 500 permutations, the estimated \textit{p}-value with \eqref{eq:pvalue} is 0.42 and we fail to reject the null hypothesis \eqref{eq:null0}, indicating that we have insufficient evidence to prove that age has an effect on the estimated covariance matrices while sex and diagnosis are in the model. Although the mean levels of TSH and thyroid hormones differ with age \citep{kapelari_pediatric_2008, aggarwal_thyroid_2013, biondi_normal_2013, park_age-and_2018}, the correlation between these hormones may not be affected by aging. Similarly, we apply the significance test for diagnosis and sex while controlling for the remaining two covariates, and the estimated \textit{p}-values for both tests are 0, which indicates that both diagnosis and sex, taken individually, have an effect on the covariance matrix of the four hormones. We compare the estimated correlations using the proposed method to the sample correlations computed using the whole sample, which are represented with the black dashed lines in Figure \ref{fig:thyroid}. For example, the sample correlation between TSH and T3 over all samples is -0.28 which is not close to the estimated correlation of either hypothyroidism or normal subjects. Furthermore, the estimated variances of the four hormones as a function of age, sex and diagnosis are presented in Supplementary Figure \ref{fig:thyroid_variance} in the Supplementary Material. We can see that the variances also differ with covariates. For a mean regression analysis for any of these hormones, assuming a constant variance could yield misleading results. 

The findings of this analysis suggest that there may be sex and diagnosis specific differences in the regulation of thyroid function, which could have important implications for the diagnosis and treatment of thyroid disorders in men and women. Clinicians can use this information to better understand the relationship between TSH and thyroid hormones in their patients, and to tailor their diagnostic and treatment approaches accordingly. It is known that the mean levels of TSH and thyroid hormones are different for hypothyroidism subjects compared to normal subjects. However, in Figure \ref{fig:thyroid}, we also observe that there is a difference in correlation between hypothyroidism and normal subject classes. Moreover, we see that there is a difference between genders for hypothyroidism subjects for TSH and thyroid hormone correlations, Cor(TSH, T3), Cor(TSH, TT4), Cor(TSH, FTI).

\section{Concluding remarks} \label{sec:conclusion}

In this study, we propose a nonparametric covariance regression method, using a random forest framework, for estimating the covariance matrix of a multivariate response given a set of covariates. Random forest trees are built with a new splitting rule designed to maximize the distance between the sample covariance matrix estimates of the child nodes. For a new observation, the random forest provides the set of nearest neighbour out-of-bag (OOB) observations which is used to estimate the conditional covariance matrix for that observation. We perform a simulation study to test the performance of the proposed method and compare it to the original Gaussian-based covariance regression model \texttt{covreg}. The average computational times of both methods for the simulations are presented in Supplementary Table \ref{tbl:time} in the Supplementary Material. We can see from the table that the proposed method is significantly faster than \texttt{covreg}. For the real data analysis, the computational time was 200.14 seconds. It should also be noted that \texttt{covreg} accounts for the uncertainty quantification in estimation of parameters which inevitably results in higher computational times compared to non-Bayesian methods. Furthermore, we propose a significance test to evaluate the effect of a subset of covariates while the other covariates are in the model. We investigate two particular cases: the global effect of covariates and the effect of a single covariate. We also propose a way to compute variable importance measures.

In this paper, we use the Euclidean distance between the upper triangular part of the two covariance matrices as splitting criterion. This is to avoid double counting the off-diagonal elements since covariance matrices are symmetric. However, several alternative splitting criteria are possible using other measures for computing distance between covariance matrices. We can use alternative distance metrics such as Frobenius norm, log-Euclidean, Kullback-Leibler divergence, Fisher Information metric, Bhattacharyya distance \citep{dryden_2009, costa_2015, bhattacharyya_1946}. Another possibility is to use test statistics as splitting criteria. There is a large literature on testing the equality of covariance matrices. Here are a few examples \cite{nagao_1973, schott_2001, wolf_2002, schott_2007, srivastava_2014} and an R package \cite{R-covTestR} that implements them. Finally, we could use a weighted Euclidean distance between covariance matrices as $d(\mathbf{D}, \mathbf{E}) =  \sqrt{\sum_{i=1}^{q}\sum_{j=i}^{q} w_{ij} (\mathbf{D}_{ij} - \mathbf{E}_{ij})^2} $. This allows to finely control the weight we wish to give to each element of the matrix. This way, the splitting criterion could be based only on the the variance terms or on the covariance terms, for example. Another possibility is that for the final covariance matrix estimation for a new observation, we can use sparse or robust covariance matrix estimations \citep{rousseeuw_fast_1999, bien_sparse_2011} using the nearest neighbour observations. Similarly, it is theoretically possible to use the sparse or robust covariance matrix estimations instead of the sample covariance matrix for the tree building process. However, the computational time could be a limiting factor. The proposed method can be applied to larger $\mathbf{X}$ dimensions. The computational time increases linearly with $\texttt{mtry}$ which is the number of covariates to randomly split at each node. It can also be adapted to larger $\mathbf{Y}$ dimensions, but the computational time could be a limitation for very large $\mathbf{Y}$ dimensions. Computing the sample covariance matrix has a time complexity $\mathcal{O}(nq^2)$ for $q$ response variables and we compute covariance matrix for each node split in each tree of the forest which necessitates many covariance matrix computations.

In \cite{alakus_conditional_2021}, we proposed a method, Random Forest with Canonical Correlation Analysis (RFCCA), which estimates the conditional canonical correlation between two multivariate data sets given the subject-related covariates. This method conditionally estimates a single parameter, the canonical correlation, that summarizes the strength of the dependency between two sets of variables. In this paper, we conditionally estimate the whole covariance matrix for one set of variables. Both methods use a splitting criterion that aims at maximizing the heterogeneity of the target parameter to build a forest of trees to obtain a set of local observations that is used to compute the final estimate. Hence, the general methodology in both papers is similar but the goals are different.

\section*{Funding}

This research was supported by the Natural Sciences and Engineering Research Council of Canada (NSERC) and by Fondation HEC Montr\'eal.

\newpage
\bibliography{refs}

\newpage
\appendix
\renewcommand{\thesection}{\arabic{section}}

\begin{center}
\LARGE{\textbf{Supplementary Material for \\Covariance regression with random forests}}
\end{center}
\vspace{0.5cm}

\renewcommand{\tablename}{Supplementary Table}
\makeatletter
\renewcommand{\ALG@name}{Supplementary Algorithm}
\makeatother

\renewcommand{\figurename}{Supplementary Figure}
\renewcommand{\thefigure}{\arabic{figure}}

\setcounter{figure}{0}
\setcounter{algorithm}{0}

\section{Final covariance matrix estimation}

Random forests were introduced as a way to get predictions by averaging the predictions from many decision trees. In other words, random forest uses the in-bag training observations within a terminal node to get an estimate from that tree, and then uses the average of all trees’ estimates as the final estimate for a new observation. Besides this traditional view, random forests can be seen as a way to find nearest neighbour observations that are close to the one we want to predict. For a new observation, the set of in-bag training observations that are in the same terminal nodes as the new observation forms the set of nearest neighbor observations, i.e. Bag of Observations for Prediction (BOP). We can define the BOP for a new observation $\mathbf{x}^{*}$ as
\begin{equation*}
    BOP(\mathbf{x}^{*}) = \bigcup\limits_{b=1}^{B} I_b(\mathbf{x}^{*}),
\end{equation*}
where $B$ is the number of trees and $I_b(\mathbf{x}^{*})$ is the set of in-bag observations in the same terminal node as $\mathbf{x}^{*}$ in the $b$th tree. 

In this paper, for a new observation $\mathbf{x}^{*}$, we form the set of nearest neighbour observations with the out-of-bag (OOB) observations. We can define the $BOP_{oob}$ for a new observation as
\begin{equation*}
    BOP_{oob}(\mathbf{x}^{*}) = \bigcup\limits_{b=1}^{B} O_b(\mathbf{x}^{*}),
\end{equation*}
where $O_b(\mathbf{x}^{*})$ is the set of OOB observations in the same terminal node as $\mathbf{x}^{*}$ in the $b$th tree.

For a new observation, we can estimate the final covariance matrix using the alternative ways described above. We perform a simulation study with the four DGPs described in Data generating process subsection of the main paper to compare the performance of the four alternative ways of computing the final covariance matrix listed below. 

\begin{enumerate}
    \item Average of all trees’ estimates computed with in-bag (IB) training observations
    \item Average of all trees’ estimates computed with out-of-bag (OOB) training observations
    \item $BOP(\mathbf{x}^{*})$ - BOP constructed with in-bag (IB) training observations
    \item $BOP_{oob}(\mathbf{x}^{*})$ - BOP constructed with out-of-bag (OOB) training observations
\end{enumerate}

\begin{figure}[H]
    \centerline{\includegraphics[width=\textwidth]{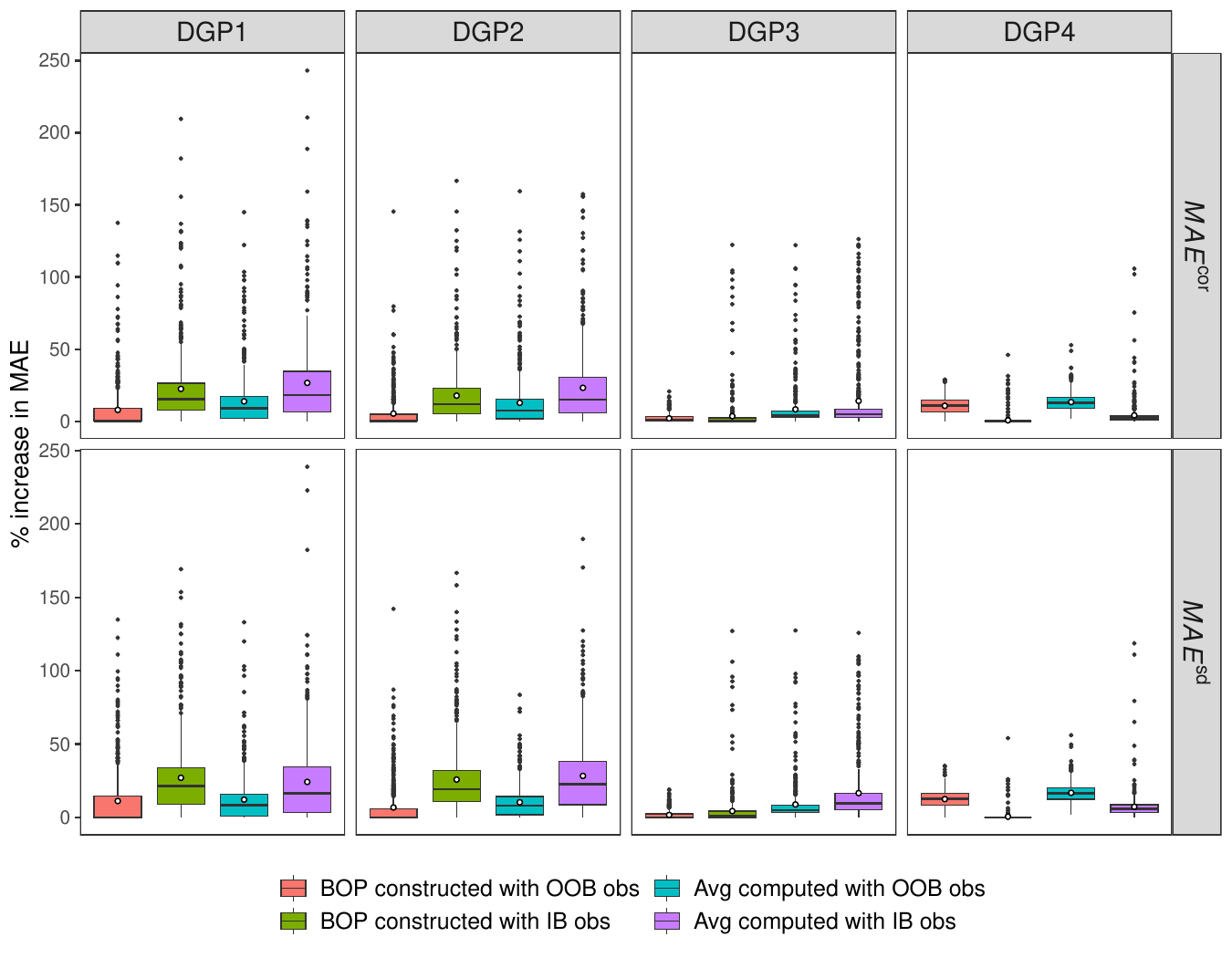}}
    \caption{Boxplots for the percentage increase in MAE of each alternative method compared to the minimum MAE for a given run across 500 runs for each DGP. The smallest is the percentage increase, the better is the method. Each white circle is the average of the relative MAE over 500 runs.}
    \label{fig:relativeacc}
\end{figure}

\begin{figure}
    \centerline{\includegraphics[width=0.6\textwidth]{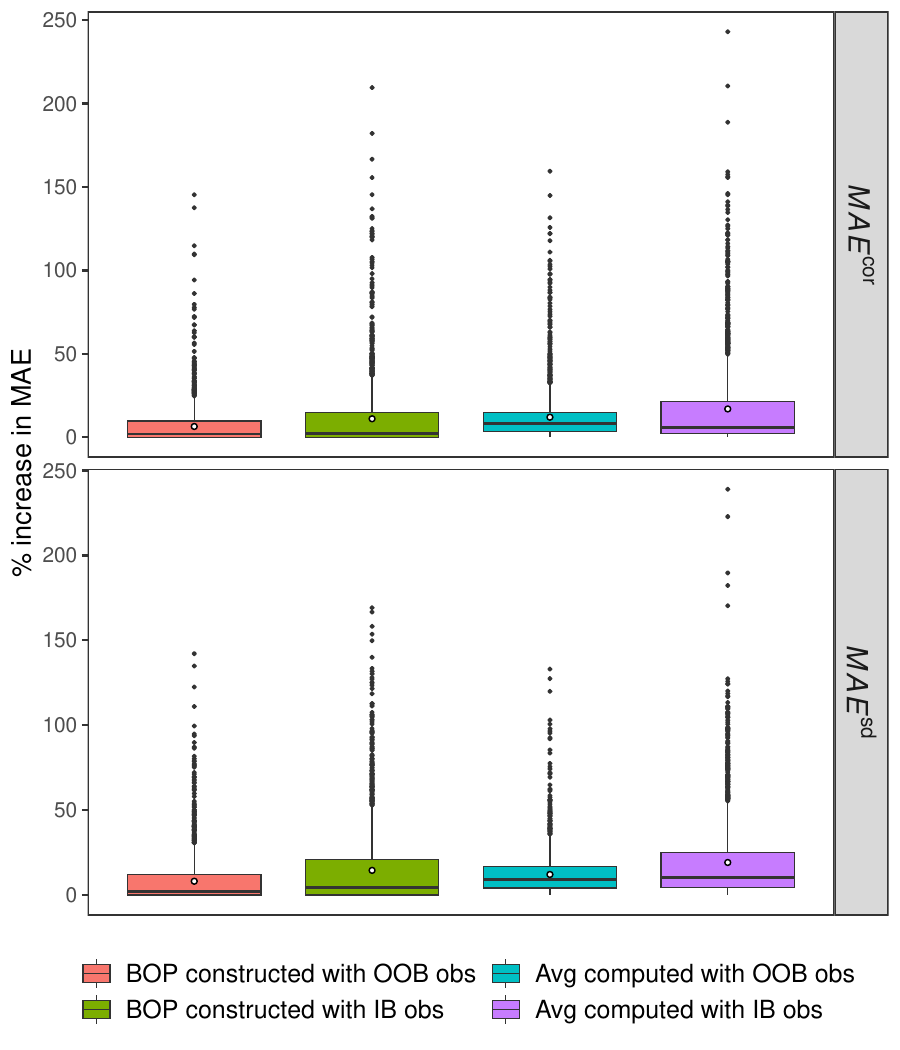}}
    \caption{Boxplots for the percentage increase in MAE of each alternative method compared to the minimum MAE for a given run across 2,000 runs for all DGPs. The smallest is the percentage increase, the better is the method. Each white circle is the average of the relative MAE over 2,000 runs.}
    \label{fig:relativeaccglobal}
\end{figure}

We can globally compare the accuracy over all scenarios with the percentage increase in MAE of a method with respect to the best method for a given run. For a given run, define $MAE_{i}$ as the mean absolute error (MAE) of method $i$ and $MAE^{*}$ as the minimum MAE over the four alternative ways of estimating the final covariance matrix. The percentage increase in MAE for method $i$ is computed as $$100 \times \frac{MAE_{i} - MAE^{*}}{MAE^{*}}.$$ Smaller values for this measure indicate better performances. Supplementary Figure \ref{fig:relativeacc} presents the relative error of the alternative ways of computing final covariance matrix across 500 runs (5 $n_{train}$ values $\times$ 100 replications) for each DGP. As demonstrated in Supplementary Figure \ref{fig:relativeacc}, for DGP1, DGP2 and DGP3, $BOP_{oob}(\mathbf{x}^{*})$ provides better accuracy compared to other three methods, whereas for DGP4, $BOP(\mathbf{x}^{*})$ provides the best accuracy. In order to compare the results globally across all DGPs, Supplementary Figure \ref{fig:relativeaccglobal} presents the relative error of the four alternative ways across 2,000 runs (4 DGPs $\times$ 5 $n_{train}$ values $\times$ 100 replications). Globally, $BOP_{oob}(\mathbf{x}^{*})$ provides slightly better accuracy among the the four alternative methods.

\newpage

\section{\texttt{nodesize} tuning}

\begin{figure}
    \centerline{\includegraphics[width=\textwidth]{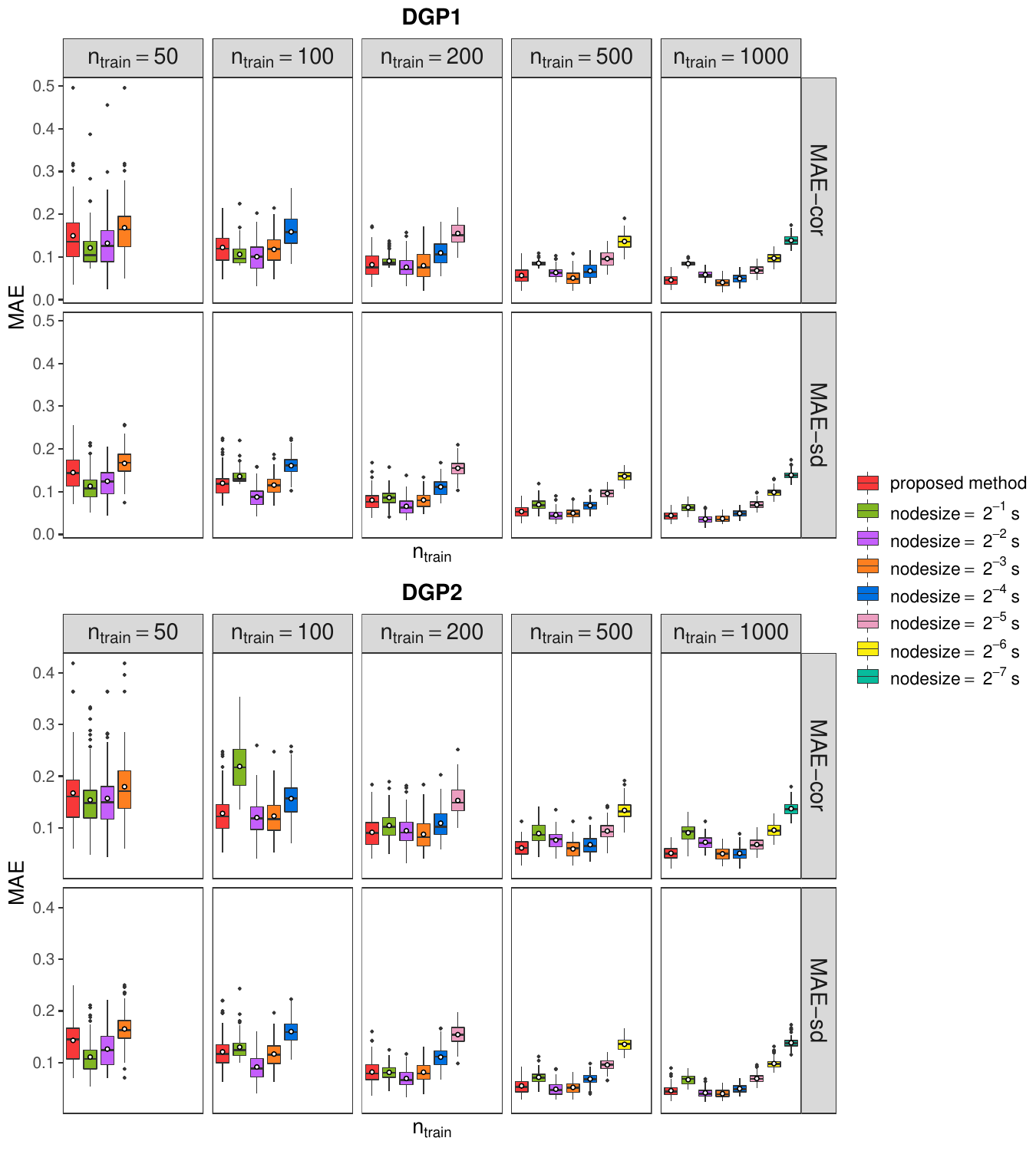}}
    \caption{MAE results for different \texttt{nodesize} values for DGP1 and DGP2. Smaller values of $MAE^{cor}$ and $MAE^{sd}$ are better. $s$ is the sub-sample size, \textit{i.e.} $s=.632 n_{train}$. Red boxplots illustrate the accuracy for the proposed \texttt{nodesize} tuning, and the rest of the boxplots show the accuracy obtained when we set the \texttt{nodesize} to a specific value.}
    \label{fig:nodesize1}
\end{figure}

\begin{figure}
    \centerline{\includegraphics[width=\textwidth]{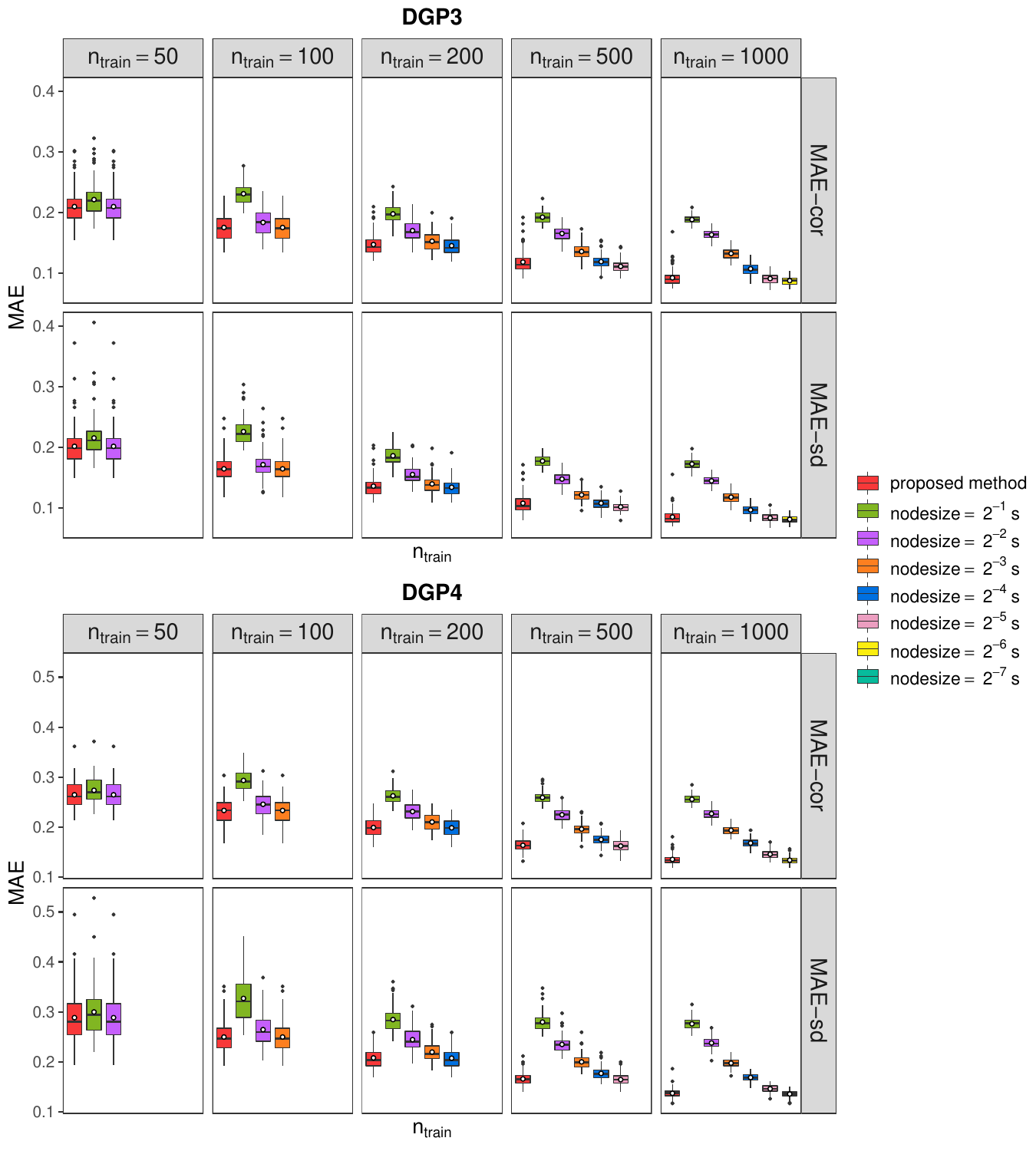}}
    \caption{MAE results for different \texttt{nodesize} values for DGP3 and DGP4. Smaller values of $MAE^{cor}$ and $MAE^{sd}$ are better. $s$ is the sub-sample size, \textit{i.e.} $s=.632 n_{train}$. Red boxplots illustrate the accuracy for the proposed \texttt{nodesize} tuning, and the rest of the boxplots show the accuracy obtained when we set the \texttt{nodesize} to a specific value.}
    \label{fig:nodesize2}
\end{figure}

Supplementary figures \ref{fig:nodesize1} and \ref{fig:nodesize2} present the accuracy results for different levels of \texttt{nodesize} along with the proposed method which applies a \texttt{nodesize} tuning as described in the main paper. In the figures, the red boxplots illustrate the MAE results for the proposed \texttt{nodesize} tuning heuristic, and the remaining boxplots show the accuracy obtained when we set the \texttt{nodesize} to a specific value. For the set of \texttt{nodesize} levels to be searched in the proposed method, we have \texttt{nodesize}$=\{[(2^{-1},2^{-2},2^{-3},\ldots)s]>q\}$ where $q$ is the number of outcomes and $s$ is the sub-sample size computed as $s=0.632n_{train}$. As can be seen from the results in Supplementary Figure \ref{fig:nodesize1}, as \texttt{nodesize} decreases, first $MAE^{cor}$ and $MAE^{sd}$ decrease and after a point increase for both DGP1 and DGP2. Since we have more levels of \texttt{nodesize} in the larger sample scenarios, it is easier to see this behaviour. For these two DGPs, smaller \texttt{nodesize} values do not mean better performance. As can be seen from Supplementary Figure \ref{fig:nodesize2}, for DGP3 and DGP4, contrary to results of DGP1 and DGP2, $MAE^{cor}$ and $MAE^{sd}$ decrease as the \texttt{nodesize} increases. Hence, the best performing \texttt{nodesize} is mostly the smallest. Overall, when we compare the accuracy of the proposed \texttt{nodesize} tuning heuristic to the individual results of different \texttt{nodesize} values, we can see that it mostly performs well, especially for the larger sample sizes.

\subsection{An example} \label{suppsec:2_1}

As an example, we illustrate the steps of the \texttt{nodesize} tuning process for DGP2 with $n_{train} = 1000$. Let $s(1) < \ldots < s(M)$ be a set of increasing node sizes and are found as \texttt{nodesize}$=\{[(2^{-1},2^{-2},2^{-3},\ldots)s]>q\}$ where $s = 0.632*1000 = 632$ and $q=2$. Therefore, \texttt{nodesize}$=\{5, 10, 20, 40, 79, 158, 316\}$. After training separate random forests for this set of \texttt{nodesize} values (7 random forests), we compute the OOB covariance matrix estimates for each forest. Let $\mathbf{\hat \Sigma}^s_{\mathbf{x}_{i}}$ be the estimated covariance matrix for observation $i$ when \texttt{nodesize}$=s$. Then, as described in the main paper, we compute $$MAD_j=\frac{1}{632}\sum_{i=1}^{632} MAD \left( \mathbf{\hat \Sigma}_{\mathbf{x}_{i}}^{s(j)},\mathbf{\hat \Sigma}_{\mathbf{x}_{i}}^{s(j+1)} \right), \ \ j =\{1, 2, \ldots, 6\}.$$ In this example, $MAD_j$ values are presented in Supplementary Table \ref{tbl:exmad}. The smallest value is $MAD_5$ which is computed with $\mathbf{\hat \Sigma}_{\mathbf{x}_{i}}^{s(5)}$ and $\mathbf{\hat \Sigma}_{\mathbf{x}_{i}}^{s(6)}$. Therefore, the best \texttt{nodesize} is $s(5)=79$.

\begin{table}[h]
\centering
\caption{$MAD_j$ values for the example. The smallest value is bold.}
\label{tbl:exmad}
\begin{tabular}{lr}
\hline
$MAD_1$ & 0.125                                 \\ \hline
$MAD_2$ & 0.099                                 \\ \hline
$MAD_3$ & 0.076                                 \\ \hline
$MAD_4$ & 0.053                                  \\ \hline
$MAD_5$ & {\color[HTML]{333333} \textbf{0.048}} \\ \hline
$MAD_6$ & 0.131                                 \\ \hline
\end{tabular}
\end{table}

\newpage
\section{Global significance test}

The proposed global significance test is described in Supplementary Algorithm \ref{alg:global}. After computing the unconditional and conditional covariance matrices, $\mathbf{\Sigma}_{root}$ and $\mathbf{\Sigma}_{\mathbf{x}_i}$, respectively, we compute the global test statistic with
\begin{equation} \label{eq:globaltest}
    T = \frac{1}{n} \sum_{i=1}^{n}{d \big(\mathbf{\hat \Sigma}_{\mathbf{x}_i}, \mathbf{\Sigma}_{root}\big)},
\end{equation}
where $d(.,.)$ is computed as (2) in the main paper.

\begin{algorithm}[]
	\caption{Global permutation test for covariates' effects} 
	\begin{algorithmic}[1]
	\State Compute sample covariance matrix of $\mathbf{Y}$ in the root node, say $\mathbf{\Sigma}_{root}$
	\State Train a RF with $\mathbf{X}$ and $\mathbf{Y}$
	\State Estimate covariance matrices as described in Algorithm 1 of the main paper, say $\mathbf{\hat \Sigma}_{\mathbf{x}_i}$ $\forall i=\{1,\ldots,n\}$
	\State Compute test statistic $T$ as in \eqref{eq:globaltest}
	\For {$r = 1:R$}
	    \State Permute rows of $\mathbf{X}$ to obtain $\mathbf{X}_{r}$
	    \State Train a RF with $\mathbf{X}_{r}$ and $\mathbf{Y}$
        \State Estimate covariance matrices as described in Algorithm 1 of the main paper, say $\mathbf{\hat \Sigma}^{'}_{\mathbf{x}_i}$ $\forall i=\{1,\ldots,n\}$
        \State Compute test statistic with $T'_r = \frac{1}{n} \sum_{i=1}^{n}{d \big(\mathbf{\hat \Sigma}^{'}_{\mathbf{x}_i}, \mathbf{\Sigma}_{root}\big)}$
    \EndFor
    \State Approximate the permutation $p$-value with $p = \frac{1}{R} \sum_{r=1}^{R}{I(T'_r > T)}$
    \State Reject the null hypothesis at level $\alpha$ when $p < \alpha$. Otherwise, do not reject the null hypothesis.
	\end{algorithmic} 
	\label{alg:global}
\end{algorithm}

\section{Variable importance computation}

\begin{algorithm}[]
	\caption{Variable importance computation} 
	\begin{algorithmic}[1]
	\State For original covariates $\mathbf{X}$ and responses $\mathbf{Y}$, estimate covariance matrices with the proposed method as described in Algorithm 1 in the main paper, say $\mathbf{\hat \Sigma}^{}_{\mathbf{x}_i}$ $\forall i=\{1,\ldots,n\}$
    \State Train a RF that uses a multivariate splitting rule based on the Mahalanobis distance with original covariates $\mathbf{X}$ to predict
    \[
    \begin{bmatrix} 
    \hat \sigma_{111} & \hat \sigma_{112} & \dots & \hat \sigma_{1qq}\\
    \hat \sigma_{211} & \hat \sigma_{212} & \dots & \hat \sigma_{2qq}\\
    \vdots & \vdots & \ddots & \vdots\\
    \hat \sigma_{n11} & \hat \sigma_{n12} & \dots & \hat \sigma_{nqq}\\
    \end{bmatrix}_{n\times \frac{q\left(q+1\right)}{2}}
    \]
    where row $i$ represents the upper triangular part of the estimated covariance matrix of observation $i$, $\hat \sigma_{ijk}$ represent the covariances in row $j$ and column $k$ of $\mathbf{\hat \Sigma}^{}_{\mathbf{x}_i}$ $i=\{1,\ldots,n\}$, $j=\{1,\ldots,q\}$, $k=\{j,\ldots,q\}$.
    \State Get the variable importance measures from this RF
	\end{algorithmic} 
	\label{alg:vimp}
\end{algorithm}

\section{Data generating process} \label{suppsec:dgp}

In DGP1, the covariance matrix for the observation $x_i$ is $$\Sigma_{\mathbf{x_i}} = \mathbf{\Psi} + \mathbf{B}\mathbf{x}_i\mathbf{x}_i^T\mathbf{B}^T,$$ where $\mathbf{x}_i^T=(1, x_i)^T$, $\mathbf{B}_0 = [(1,-1)^T,(1, 1)^T]$, $\mathbf{B} = \frac{w}{w+1}\mathbf{B}_0 $, $\mathbf{\Psi}_0 = \mathbf{B}_0 [(1,0)^T,(0, 1/3)^T] \mathbf{B}^T_0$, $\mathbf{\Psi} = \frac{1}{w+1}\mathbf{\Psi}_0 $ and $w=1$.

In DGP3, the correlations are generated with all seven covariates according to a tree model with a depth of three and eight terminal nodes:
\begin{align*}
\rho(\mathbf{x}_i) & = u_1 I\left(x_{i1}<0, x_{i2}<0, x_{i4}<0\right)\\
& + u_2 I\left(x_{i1}<0, x_{i2}<0, x_{i4} \geq 0\right)\\
& + u_3 I\left(x_{i1}<0, x_{i2} \geq 0, x_{i5}<0\right)\\
& + u_4 I\left(x_{i1}<0, x_{i2} \geq 0, x_{i5} \geq 0\right)\\
& + u_5 I\left(x_{i1} \geq 0, x_{i3}<0, x_{i6}<0\right)\\
& + u_6 I\left(x_{i1} \geq 0, x_{i3}<0, x_{i6} \geq 0\right)\\
& + u_7 I\left(x_{i1} \geq 0, x_{i3} \geq 0, x_{i7}<0\right)\\
& + u_8 I\left(x_{i1} \geq 0, x_{i3} \geq 0, x_{i7} \geq 0\right),
\end{align*}
where the terminal node values are $u=\left(0.2, 0.3, 0.4, 0.5, 0.6, 0.7, 0.8, 0.9\right)$ and $I$ is the indicator function. The variances are functions of $\rho$ and computed as $Var(y_j|\mathbf{x}_i) = (1+\rho(\mathbf{x}_i))^j$, $j=\{1,\ldots,q\}$.

In DGP4, for an observation $\mathbf{x}_i$, we can generate the correlation with the logit model, 
$$\rho(\mathbf{x}_i) = \frac{1}{1 + \exp{\big(-(\beta_0 + \sum_{j=1}^{p} \beta_j x_{ij} + x_{i1}^2)}\big)},$$ 
where $\beta_0^{}$ is the intercept parameter fixed to $\beta_0 = -1$ and $\beta_j$ are the weights for the covariates, fixed to $(1, 1-\frac{1}{p}, 1-\frac{2}{p}, \ldots, 1-\frac{(p-1)}{p})$. For an observation $\mathbf{x}_i$, the variance of each response is generated as $Var(y_j|\mathbf{x}_i) = (1+\rho(\mathbf{x}_i))^j$, $j=\{1,\ldots,q\}$. 

\section{Difference in MAE between \texttt{CovRegRF} and competing methods}

\begin{figure}[!h]
    \centerline{\includegraphics[width=0.7\textwidth]{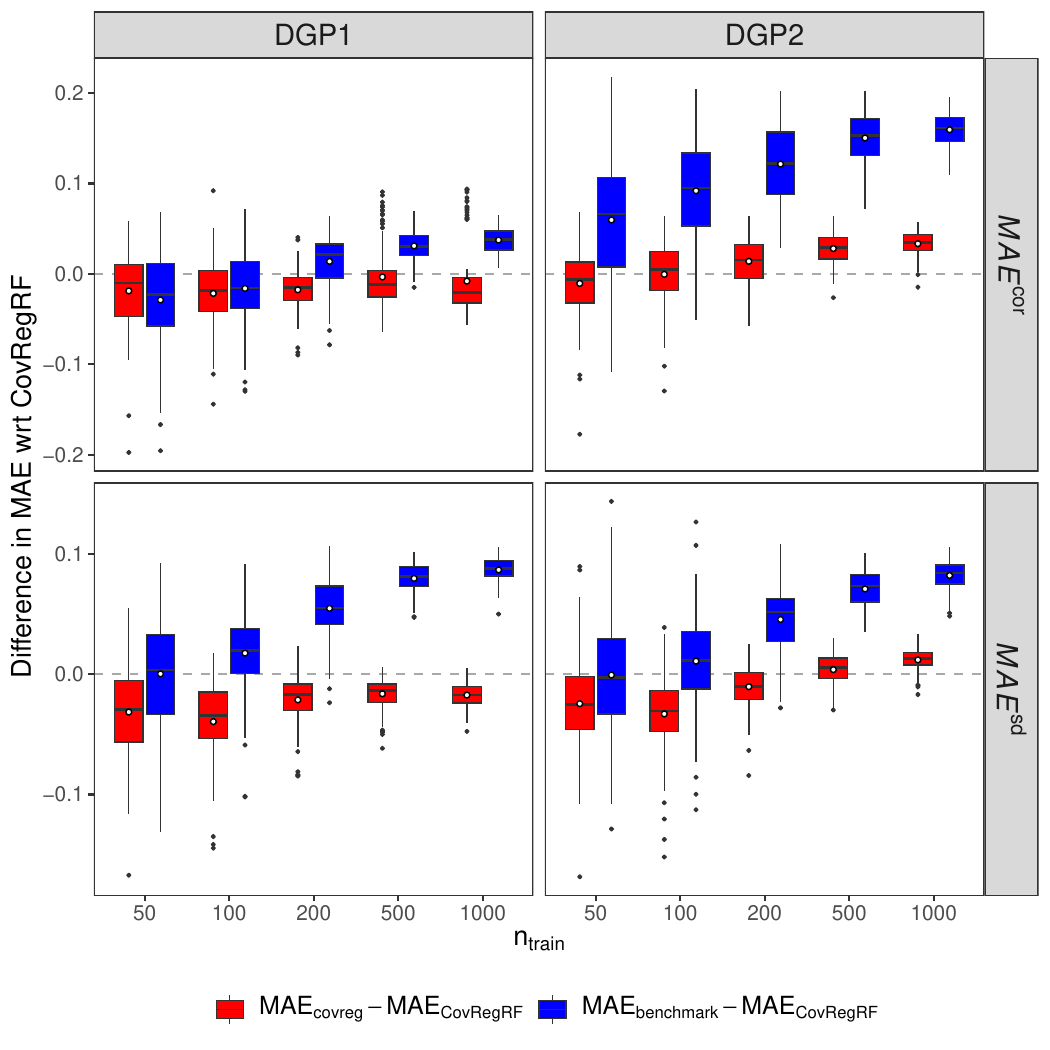}}
    \caption{Difference in MAE between \textit{(red)} \texttt{covreg} and \texttt{CovRegRF}, and \textit{(blue)} benchmark and \texttt{CovRegRF}. The dashed line is at 0.}
    \label{fig:acc1}
\end{figure}

\begin{figure}[!h]
    \centerline{\includegraphics[width=0.7\textwidth]{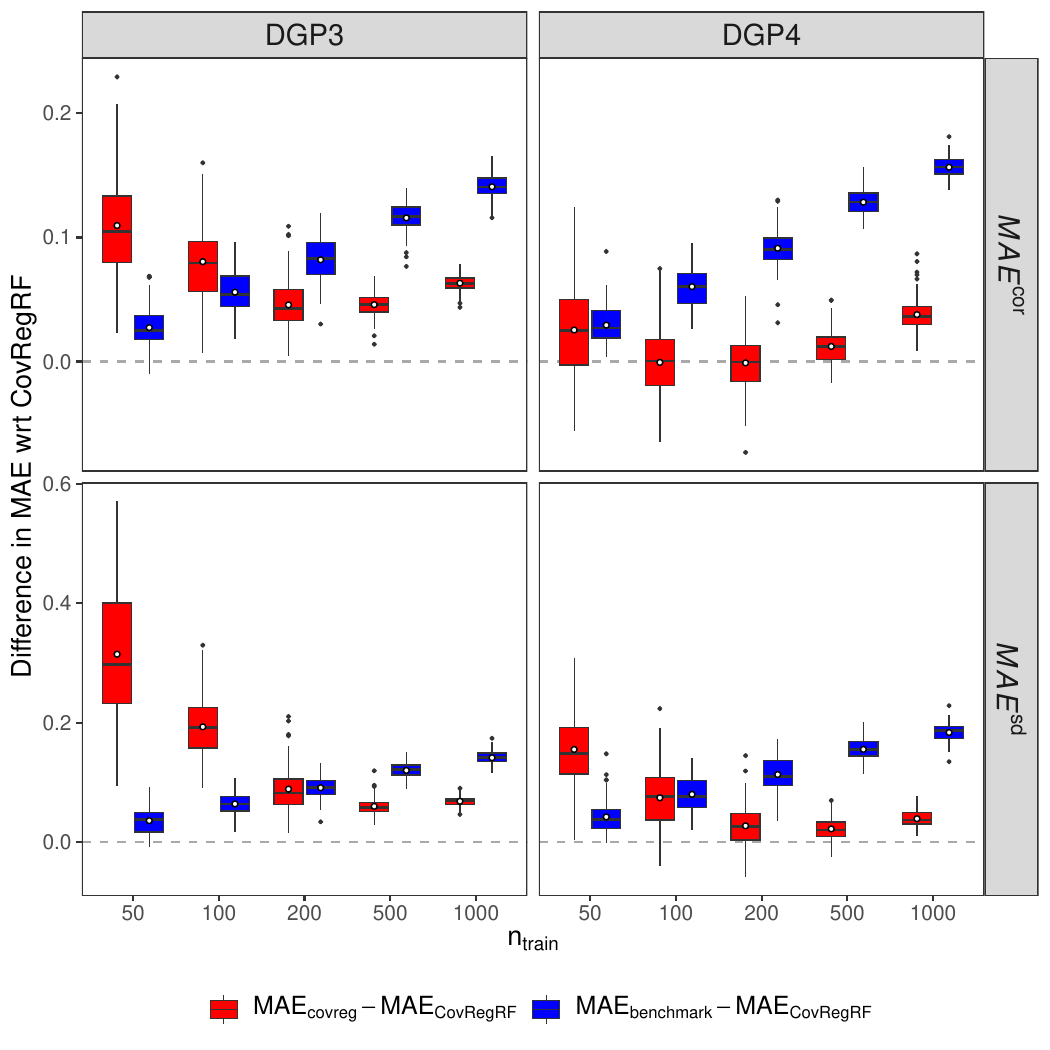}}
    \caption{Difference in MAE between \textit{(red)} \texttt{covreg} and \texttt{CovRegRF}, and \textit{(blue)} benchmark and \texttt{CovRegRF}. The dashed line is at 0.}
    \label{fig:acc2}
\end{figure}

Supplementary figures \ref{fig:acc1} and \ref{fig:acc2} present the difference in MAE between \textit{(red boxplots)} \texttt{covreg} and \texttt{CovRegRF}, and \textit{(blue boxplots)} benchmark and \texttt{CovRegRF} results for 100 repetitions. In the boxplots, the values greater than 0 (above the dashed line) demonstrate that \texttt{CovRegRF} has smaller MAE than the competing method. On the contrary, the values less than 0 show that \texttt{CovRegRF} has larger MAE than the competing method. 

\section{Accuracy evaluation with Stein's loss}

\begin{figure}[!h]
    \centerline{\includegraphics[width=0.9\textwidth]{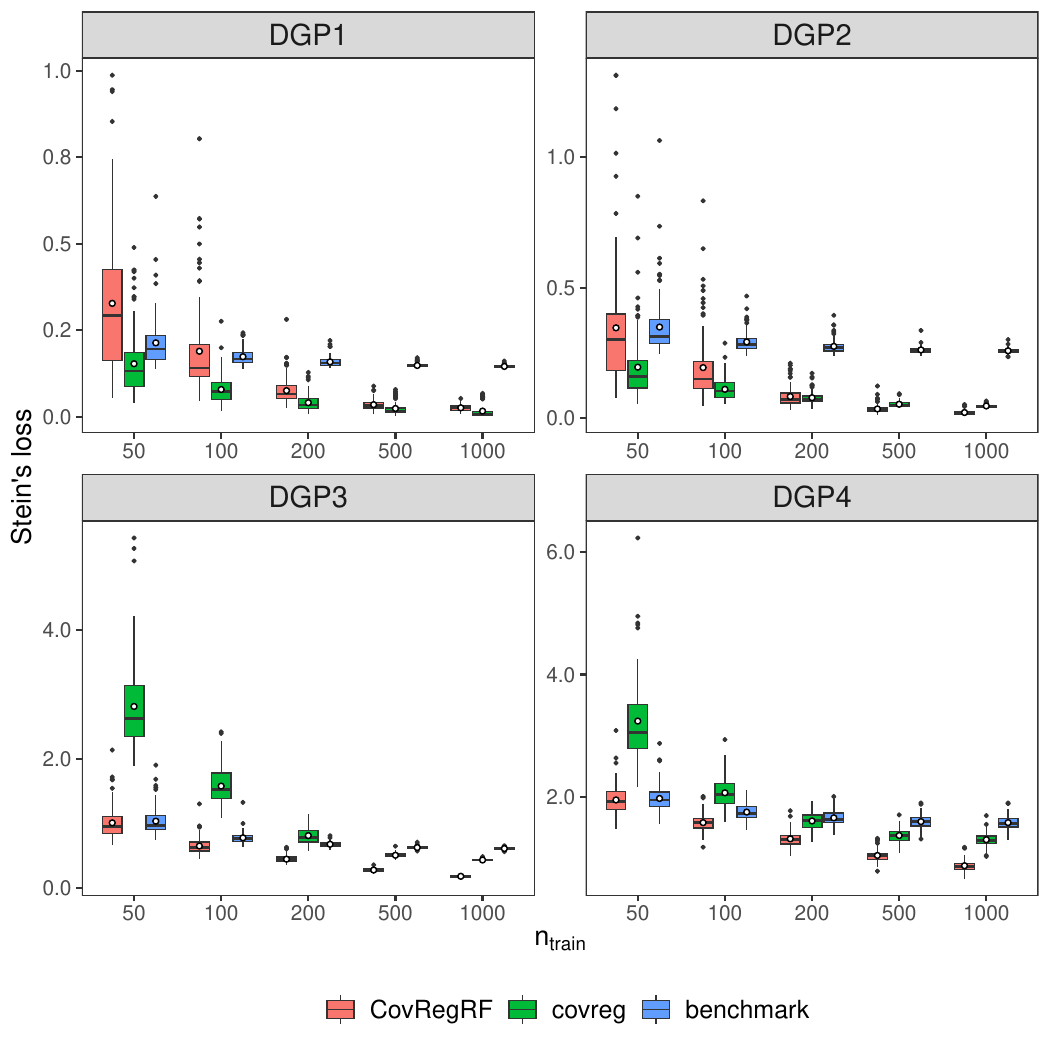}}
    \caption{Accuracy evaluation for four DGPs with Stein's loss. Smaller values of this metric are better.}
    \label{fig:acc3}
\end{figure}

In addition to mean absolute errors (MAE) computed for the estimated correlations and standard deviations, we can compare each estimated covariance matrix $\mathbf{\hat \Sigma}_{\mathbf{x}_i}$ to its corresponding true matrix $\mathbf{\Sigma}_{\mathbf{x}_i}$ with Stein's loss which is the Kullback–Leibler divergence between two multivariate normal distributions with means zero and covariance matrices $\mathbf{\hat \Sigma}_{\mathbf{X}}$ and $\mathbf{\Sigma}_{\mathbf{X}}$,

\begin{equation*}
    l(\mathbf{\hat \Sigma}_{\mathbf{X}}, \mathbf{\Sigma}_{\mathbf{X}}) = \frac{1}{n_{test}}\sum_{i=1}^{n_{test}} \left ( \Tr \left (\mathbf{\hat \Sigma}_{\mathbf{x}_i}^{-1} \mathbf{\Sigma}_{\mathbf{x}_i}^{} \right) - \log  \det \left(\mathbf{\hat \Sigma}_{\mathbf{x}_i}^{-1} \mathbf{\Sigma}_{\mathbf{x}_i}^{} \right) - q  \right)
\end{equation*}
where $q$ is the number of responses. Each term in the sum becomes 0 when $\mathbf{\hat \Sigma}_{\mathbf{x}_i} = \mathbf{\Sigma}_{\mathbf{x}_i}$. Therefore, smaller Stein's loss values correspond to better covariance matrix estimates. Supplementary Figure \ref{fig:acc3} presents the Stein's loss for \texttt{CovRegRF}, \texttt{covreg} and benchmark for all DGPs for 100 repetitions. 

For all DGPs, with increasing sample size, both the proposed method and \texttt{covreg} improve and the variance in Stein's loss decreases. For DGP1, \texttt{covreg} performs better compared to the proposed method for all sample sizes. However, the difference between \texttt{covreg} and the proposed method decreases with increasing sample size. For DGP2, for smaller sample sizes, \texttt{covreg} performs better whereas after $n_{train}=500$, the proposed method performs slightly better. For DGP3, the proposed method performs significantly better than \texttt{covreg} for all sample sizes. Similarly, for DGP4, the proposed method has smaller Stein's loss compared to \texttt{covreg} for all sample sizes.

\section{Simulation results for variable importance}

\begin{figure}[!h]
    \centerline{\includegraphics[width=0.7\textwidth]{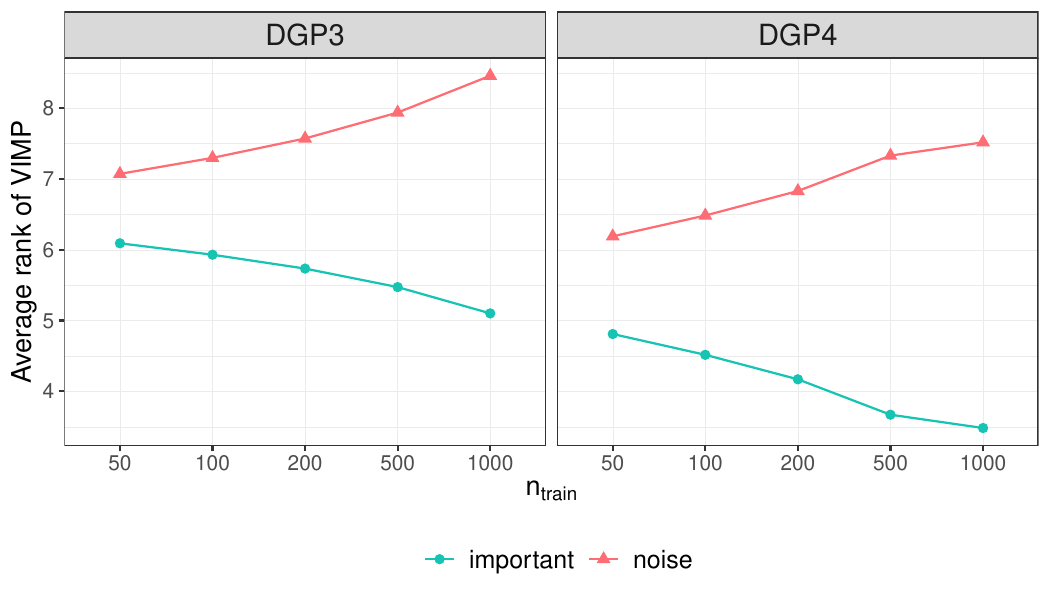}}
    \caption{Average ranks from estimated VIMP measures for DGP3 and DGP4. Smaller rank values indicate a more important variable (the most important variable has rank 1).}
    \label{fig:vimp}
\end{figure}

As stated in the main paper, Supplementary Figure \ref{fig:vimp} presents the average rank, from the estimated VIMP measures, for the important and noise variables groups for DGP3 and DGP4. The variable with the highest VIMP measure has rank 1. As rank increases, variable importance decreases. In all scenarios, the important variables have smaller average ranks than noise variables. As expected, the difference between the average ranks of important and noise variables increases with increasing sample size.

\section{Real data example}

\begin{figure}[!h]
    \centerline{\includegraphics[width=0.7\textwidth]{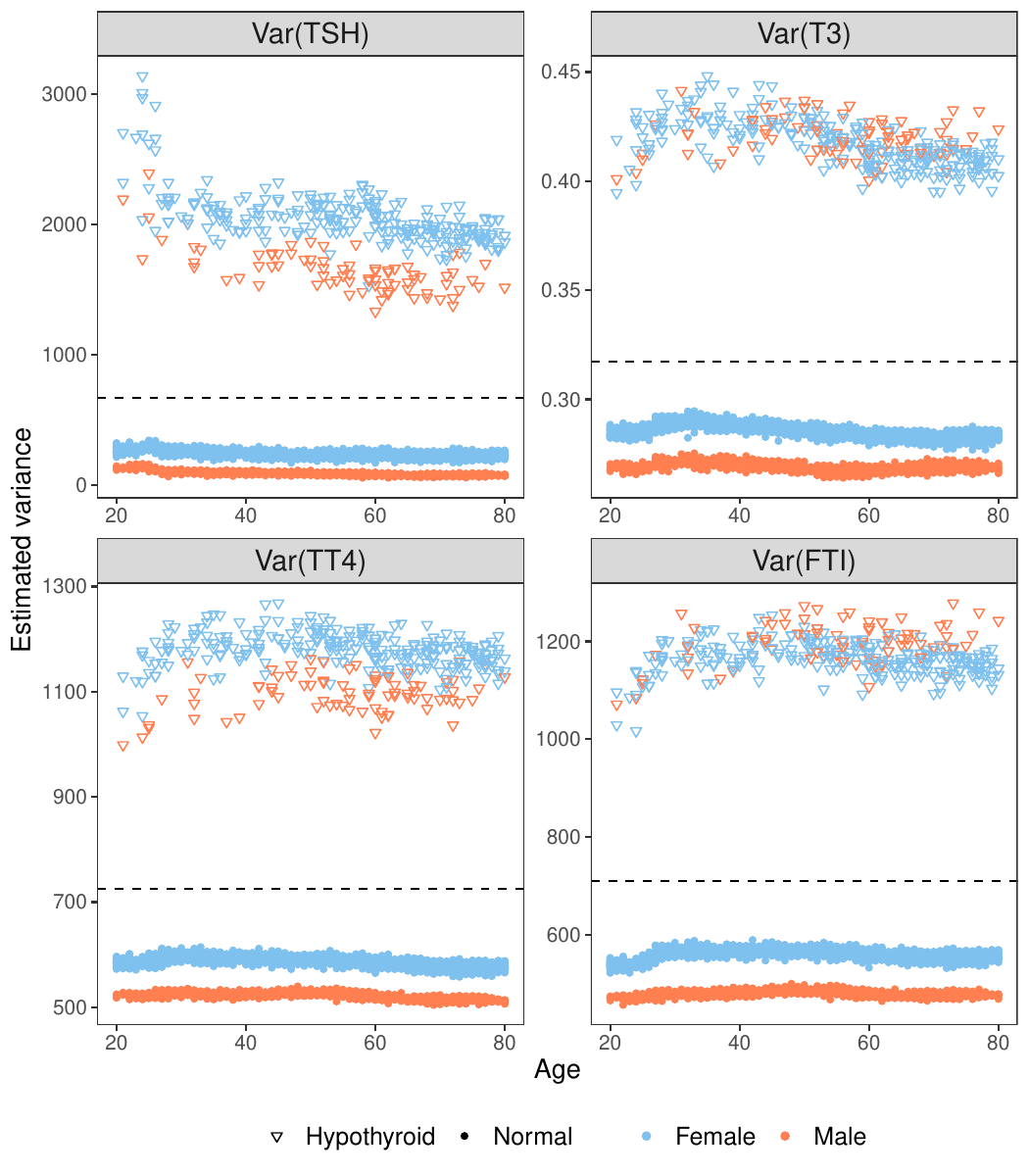}}
    \caption{Estimated variances for the four hormones as a function of age, sex and diagnosis. Dashed lines represent the sample variances computed using the whole sample.}
    \label{fig:thyroid_variance}
\end{figure}

Supplementary Figure \ref{fig:thyroid_variance} presents the estimated variances of the four hormones (TSH, T3, TT4 and FTI) as a function of age, sex, and diagnosis. As we can see from the plots, the variances differ with diagnosis and sex, whereas age does not seem to have much effect on the estimated variances. 

\section{Comparison of computational times}

All simulations were run in R version 3.6.0 on a Linux machine with Intel(R) Xeon(R) E5-2667 v3 @ 3.20GHz with 396 GB of memory. The average computational time of each method for the four DGPs is presented in Supplementary Table \ref{tbl:time}. For both methods, the time for a setting consists of the time for training and the time for prediction for a new data set. We can see that the proposed method is significantly faster than \texttt{covreg}.

\begin{table}[h]
\caption{Average computational time (in seconds) of both methods over 100 replications for each simulated data set.}
\label{tbl:time}
\centering
\begin{tabular}[t]{rlrr}
\toprule
$n_{train}$ & DGP & \texttt{CovRegRF} & \texttt{covreg}\\
\midrule
 & DGP1 & 2.89 & 148.23\\

 & DGP2 & 2.85 & 149.12\\

 & DGP3 & 2.60 & 304.43\\

\multirow[t]{-4}{*}{\raggedleft\arraybackslash 50} & DGP4 & 2.46 & 248.27\\
\cmidrule{1-4}
 & DGP1 & 4.01 & 151.55\\

 & DGP2 & 4.01 & 151.97\\

 & DGP3 & 3.74 & 283.25\\

\multirow[t]{-4}{*}{\raggedleft\arraybackslash 100} & DGP4 & 3.44 & 247.77\\
\cmidrule{1-4}
 & DGP1 & 6.46 & 228.57\\

 & DGP2 & 6.67 & 229.55\\

 & DGP3 & 6.48 & 428.89\\

\multirow[t]{-4}{*}{\raggedleft\arraybackslash 200} & DGP4 & 5.82 & 495.07\\
\cmidrule{1-4}
 & DGP1 & 15.07 & 383.16\\

 & DGP2 & 14.77 & 384.52\\

 & DGP3 & 15.38 & 593.00\\

\multirow[t]{-4}{*}{\raggedleft\arraybackslash 500} & DGP4 & 13.49 & 744.41\\
\cmidrule{1-4}
 & DGP1 & 52.64 & 771.69\\

 & DGP2 & 52.34 & 739.52\\

 & DGP3 & 62.96 & 984.28\\

\multirow[t]{-4}{*}{\raggedleft\arraybackslash 1000} & DGP4 & 53.95 & 1318.28\\
\bottomrule
\end{tabular}
\end{table}

\end{document}